\documentclass[prd,superscriptaddress,floatfix,notitlepage,reprint]{revtex4-1} 
\pdfoutput=1
\usepackage{amssymb,epsfig}
\usepackage{amsmath}
\usepackage[]{empheq}
\usepackage{hyperref}
\usepackage{natbib,ifthen}

\usepackage{wasysym}
\usepackage{mathrsfs}
\usepackage{simplewick}
\usepackage[lofdepth,lotdepth,caption=false]{subfig}
\usepackage{color}
\usepackage{bbold}

\begin{document}

\newcommand{\jcap}{JCAP}
\newcommand{\apjl}{APJL~}

\newcommand{\FFPT}{\textsf{FFT-PT}}
\def\Btheo{{B_\delta^{\textrm{theo}}}}
\newcommand{\todo}[1]{{\color{blue}{TODO: #1}}}
\newcommand{\comment}[1]{{\color{blue}{#1}}} 
\newcommand{\zvcomment}[1]{{\color{blue}{ZV: #1}}}
\newcommand{\lbar}[1]{\underline{l}_{#1}}
\newcommand{\drm}{\mathrm{d}}
\renewcommand{\d}{\mathrm{d}}
\renewcommand{\rm}[1]{\mathrm{#1}}
\newcommand{\gaensli}[1]{\lq #1\rq$ $}
\newcommand{\bartilde}[1]{\bar{\tilde #1}}
\newcommand{\barti}[1]{\bartilde{#1}}
\newcommand{\ti}{\tilde}
\newcommand{\oforder}[1]{\mathcal{O}(#1)}
\newcommand{\D}{\mathrm{D}}
\renewcommand{\(}{\left(}
\renewcommand{\)}{\right)}
\renewcommand{\[}{\left[}
\renewcommand{\]}{\right]}
\def\<{\left\langle}
\def\>{\right\rangle}
\newcommand{\mycaption}[1]{\caption{\footnotesize{#1}}}
\newcommand{\hattilde}[1]{\hat{\tilde #1}}
\newcommand{\mycite}[1]{[#1]}
\newcommand{\mnras}{Mon.\ Not.\ R.\ Astron.\ Soc.}
\newcommand{\apjs}{Astrophys.\ J.\ Supp.}

\def\uk{{\bf \hat{k}}}
\def\un{{\bf \hat{n}}}
\def\ur{{\bf \hat{r}}}
\def\ux{{\bf \hat{x}}}
\def\bk{{\bf k}}
\def\bn{{\bf n}}
\def\br{{\bf r}}
\def\bx{{\bf x}}
\def\bK{{\bf K}}
\def\by{{\bf y}}
\def\bl{{\bf l}}
\def\bkp{{\bf k^\pr}}
\def\brp{{\bf r^\pr}}

\newcommand{\fixme}[1]{{\textbf{Fixme: #1}}}
\newcommand{\detD}{{\det\!\cld}}
\newcommand{\clh}{\mathcal{H}}
\newcommand{\ud}{{\rm d}}
\renewcommand{\eprint}[1]{\href{http://arxiv.org/abs/#1}{#1}}
\newcommand{\adsurl}[1]{\href{#1}{ADS}}
\newcommand{\ISBN}[1]{\href{http://cosmologist.info/ISBN/#1}{ISBN: #1}}
\newcommand{\vort}{\varpi}
\newcommand\ba{\begin{eqnarray}}
\newcommand\ea{\end{eqnarray}}
\newcommand\be{\begin{equation}}
\newcommand\ee{\end{equation}}
\newcommand\lagrange{{\cal L}}
\newcommand\cll{{\cal L}}
\newcommand\cln{{\cal N}}
\newcommand\clx{{\cal X}}
\newcommand\clz{{\cal Z}}
\newcommand\clv{{\cal V}}
\newcommand\cld{{\cal D}}
\newcommand\clt{{\cal T}}

\newcommand\clo{{\cal O}}
\newcommand{\cla}{{\cal A}}
\newcommand{\clp}{{\cal P}}
\newcommand{\clr}{{\cal R}}
\newcommand{\uD}{{\mathrm{D}}}
\newcommand{\calE}{{\cal E}}
\newcommand{\calB}{{\cal B}}
\newcommand{\curl}{\,\mbox{curl}\,}
\newcommand\del{\nabla}
\newcommand\Tr{{\rm Tr}}
\newcommand\half{{\frac{1}{2}}}
\newcommand\fourth{{1\over 8}}
\newcommand\bibi{\bibitem}
\newcommand{\kf}{\beta}
\newcommand{\kfprod}{\alpha}
\newcommand\calS{{\cal S}}
\renewcommand\H{{\cal H}}
\newcommand\K{{\rm K}}
\newcommand\mK{{\rm mK}}
\newcommand\synch{\text{syn}}
\newcommand\opacity{\tau_c^{-1}}

\newcommand{\Psil}{\Psi_l}
\newcommand{\bsigma}{{\bar{\sigma}}}
\newcommand{\bI}{\bar{I}}
\newcommand{\bq}{\bar{q}}
\newcommand{\bv}{\bar{v}}
\renewcommand\P{{\cal P}}
\newcommand{\numfrac}[2]{{\textstyle \frac{#1}{#2}}}

\newcommand{\la}{\langle}
\newcommand{\ra}{\rangle}
\newcommand{\lla}{\left\langle}
\newcommand{\rra}{\right\rangle}

\newcommand{\vnabla}{\ensuremath{\boldsymbol\nabla}}

\newcommand{\Omtot}{\Omega_{\mathrm{tot}}}
\newcommand\xx{\mbox{\boldmath $x$}}
\newcommand{\phpr} {\phi'}
\newcommand{\gam}{\gamma_{ij}}
\newcommand{\sqgam}{\sqrt{\gamma}}
\newcommand{\delk}{\Delta+3{\K}}
\newcommand{\dph}{\delta\phi}
\newcommand{\om} {\Omega}
\newcommand{\dom}{\delta^{(3)}\left(\Omega\right)}
\newcommand{\rar}{\rightarrow}
\newcommand{\Rar}{\Rightarrow}
\newcommand\gsim{ \lower .75ex \hbox{$\sim$} \llap{\raise .27ex \hbox{$>$}} }
\newcommand\lsim{ \lower .75ex \hbox{$\sim$} \llap{\raise .27ex \hbox{$<$}} }
\newcommand\bigdot[1] {\stackrel{\mbox{{\huge .}}}{#1}}
\newcommand\bigddot[1] {\stackrel{\mbox{{\huge ..}}}{#1}}
\newcommand{\Mpc}{\text{Mpc}}
\newcommand{\Al}{{A_l}}
\newcommand{\Bl}{{B_l}}
\newcommand{\eAl}{e^\Al}
\newcommand{\ix}{{(i)}}
\newcommand{\ixp}{{(i+1)}}
\renewcommand{\k}{\beta}
\newcommand{\HD}{\mathrm{D}}

\newcommand{\nonflat}[1]{#1}
\newcommand{\Cgl}{C_{\text{gl}}}
\newcommand{\Cgltwo}{C_{\text{gl},2}}
\newcommand{\He}{{\rm{He}}}
\newcommand{\Mhz}{{\rm MHz}}
\newcommand{\vx}{{\mathbf{x}}}
\newcommand{\ve}{{\mathbf{e}}}
\newcommand{\vv}{{\mathbf{v}}}
\newcommand{\vk}{{\mathbf{k}}}
\renewcommand{\vr}{{\mathbf{r}}}
\newcommand{\vn}{{\mathbf{n}}}
\newcommand{\vPsi}{{\mathbf{\Psi}}}
\newcommand{\vs}{{\mathbf{s}}}
\newcommand{\vH}{{\mathbf{H}}}
\newcommand{\theo}{\mathrm{th}}
\newcommand{\sgn}{\mathrm{sgn}}

\newcommand{\vnhat}{{\hat{\mathbf{n}}}}
\newcommand{\vkhat}{{\hat{\mathbf{k}}}}
\newcommand{\taueps}{{\tau_\epsilon}}
\newcommand{\valpha}{\ensuremath{\boldsymbol\alpha}}
\newcommand{\vbeta}{\ensuremath{\boldsymbol\beta}}

\newcommand{\vgrad}{{\mathbf{\nabla}}}
\newcommand{\fbarln}{\bar{f}_{,\ln\epsilon}(\epsilon)}

\newcommand{\secref}[1]{Section \ref{se:#1}}
\newcommand{\expt}{\mathrm{expt}}
\newcommand{\eq}[1]{(\ref{eq:#1})} 
\newcommand{\eqq}[1]{Eq.~(\ref{eq:#1})} 
\newcommand{\fig}[1]{Fig.~\ref{fig:#1}} 
\renewcommand{\to}{\rightarrow}
\renewcommand{\(}{\left(}
\renewcommand{\)}{\right)}
\renewcommand{\[}{\left[}
\renewcommand{\]}{\right]}
\renewcommand{\vec}[1]{\mathbf{#1}}
\newcommand{\vy}{\vec{y}}
\newcommand{\vz}{\vec{z}}
\newcommand{\vq}{\vec{q}}
\newcommand{\vp}{\vec{p}}
\newcommand{\va}{\vec{a}}
\newcommand{\vb}{\vec{b}}
\newcommand{\VPsi}{\vec{\Psi}}
\newcommand{\vecv}{\vec{v}}
\newcommand{\vl}{{\boldsymbol \ell}}
\newcommand{\vell}{\boldsymbol{\ell}}
\newcommand{\VL}{\vec{L}}
\newcommand{\dl}{\d^2\vl}
\renewcommand{\L}{\mathscr{L}}

\newcommand{\abs}[1]{\lvert #1\rvert}

\newcommand{\ul}{\underline{l}}
\newcommand{\lin}{\mathrm{lin}}

\newcommand*{\df}  {\delta}
\newcommand*{\tf}  {\theta}
\renewcommand{\vec}{\textbf}
\newcommand*{\non}  {\nonumber}
\newcommand*{\lb}  {\left(}
\newcommand*{\rb}  {\right)}
\newcommand*{\ls}  {\left[}
\newcommand*{\rs}  {\right]}

\def\pvm#1{{\color{blue}[PM: {\it #1}] }}
\def\zv#1{{\color{red}[ZV: {\it #1}] }}


\thispagestyle{empty}

\title{FFT-PT: Reducing the two-loop large-scale structure power spectrum\newline
to low-dimensional radial integrals}

\author{Marcel Schmittfull}
\affiliation{Berkeley Center for Cosmological Physics, Department of
  Physics and Lawrence Berkeley
  National Laboratory, University of California, Berkeley, CA 94720, USA}

\author{Zvonimir Vlah}
\affiliation{Stanford Institute for Theoretical Physics and Department of Physics, Stanford University, Stanford, CA 94306, USA}
\affiliation{Kavli Institute for Particle Astrophysics and Cosmology, SLAC and Stanford University, Menlo Park, CA 94025, USA}

\date{\today}

\begin{abstract}

Modeling the large-scale structure of the universe on nonlinear scales has the potential to substantially increase the science return of upcoming surveys by increasing the number of modes available for model comparisons. 
One way to achieve this is to model nonlinear scales perturbatively.
Unfortunately, this involves high-dimensional 
loop integrals that are cumbersome to evaluate. 
Trying to simplify this, we show how two-loop (next-to-next-to-leading order) corrections to the
density power spectrum can be reduced to low-dimensional, radial integrals.
Many of those can be evaluated with a one-dimensional 
Fast Fourier Transform, which is
significantly faster than the five-dimensional Monte-Carlo 
integrals that are needed otherwise.
The general idea of this \textsf{FFT-PT} method is to switch
between Fourier and position space to avoid convolutions
and integrate over orientations, leaving only radial integrals.
This reformulation is independent of the underlying shape of the
initial linear density power spectrum and should easily accommodate
features such as those from baryonic acoustic oscillations.
We also discuss how to account for halo bias and redshift space 
distortions. 

\end{abstract}

\maketitle

\section{Introduction}

Observations of the large-scale structure (LSS) of the universe are becoming increasingly precise and abundant, with many large surveys planned in the near future, including e.g.~DES \cite{DESwhitepaper}, eBOSS \cite{eBOSSDawson}, DESI \cite{DESIwhitepaper}, Euclid \cite{EuclidWhitePaper}, WFIRST \cite{WFIRST1503}, LSST \cite{LSSTDESC}, and SPHEREx \cite{spherex1412}.
It is exciting to use this observational window to study fundamental physics and the evolution and composition of the universe. 
This is possible because properties of the constituents of the universe leave characteristic fingerprints in the observed distribution of LSS, enabling detailed studies of e.g.~dark energy, the initial conditions from the big bang, neutrino-like particles, or modifications of general relativity.
The accuracy with which we can study these fingerprints is set by the number of independent three-dimensional modes that we can model and include in data analyses.
This is in turn determined by the smallest scale that we can still model.
Therefore, an important aspect of large-scale structure research is to  extend the validity of models to smaller, more nonlinear scales.

Given the immense effort put into future surveys and the strong dependence of their science output on the smallest scale that can be modeled, any idea for improving LSS models on small scales is worth pursuing. 
This has therefore been an area of intense study in the literature. 
The two main perturbative modeling approaches are Eulerian standard perturbation theory (SPT) (e.g.~\cite{Goroff:1986ep, Jain:1993jh, 1996ApJS..105...37S, 1996ApJ...473..620S, Blas:2013bpa}) and Lagrangian perturbation theory (LPT) (e.g.~\cite{Zeldovich:1969sb, Bouchet:1994xp, Matsubara:2007wj}); see~\cite{bernardeauReview} for a review and e.g.~\cite{2006PhRvD..73f3519C,2008JCAP...10..036P,2008PhRvD..78j3521B,2009JCAP...06..017L,Taruya2010,2012JCAP...07..051B,2012JHEP...09..082C,2012JCAP...01..019P,2012PhRvD..85l3519B,2012MNRAS.427.2537C,2013JCAP...08..037P,2014PhRvD..90d3537M,2014JCAP...07..056C,2014JCAP...03..006M,2014JCAP...07..057C,2014PhRvD..90b3518C,2014JCAP...05..022P,2014JCAP...05..022P,2014JCAP...09..047M,2015JCAP...02..013S,zvonimir1410,2015PhRvD..91l3516S,2015JCAP...09..014V,2016JCAP...03..057V,2016JCAP...01..043M}
for a selection of more recent developments.
Higher-order perturbative corrections to these models push their validity to smaller scales. 
However these corrections involve high-dimensional, computationally expensive loop integrals. 
For example, the 2-loop power spectrum in SPT involves five-dimensional integrals at every wavenumber of interest.
Accurate numerical evaluation of the 2-loop power spectrum can therefore take several CPU hours for a single set of cosmological parameter values.
Reducing the computational complexity can make these 2-loop integrals more practicable for the LSS community, and simplify their use for constraining cosmological parameters from LSS surveys with Monte-Carlo chains, which often require evaluating  model predictions for thousands of cosmological parameter values.

Motivated by this, we recently proposed a fast method to evaluate the
1-loop, next-to-leading-order matter power spectrum from an arbitrary 
linear input power spectrum \cite{MarcelZvonimirPat1603}.
Ref.~\cite{OSUloops1603} presented the same method for 2-2 contributions 
and an alternative method for 1-3 couplings.
Related work that separates
high-dimensional integrals into products of lower dimensional integrals
includes \cite{Mccollough1202,sherwinZaldarriaga,fergusson1008,Marcel1108,Marcel1207,zvonimir1410,slepian1411,MarcelTobiasUros1411,Slepian1607} for LSS and e.g.~\cite{KSW,fergusson0912,kendrickTrispectrum1502,BoehmN32} for the CMB.

Our method in \cite{MarcelZvonimirPat1603} executes 20 one-dimensional FFTs to return the 1-loop power spectrum over several decades in wavenumber at once at machine-level precision.
This exploits spherical symmetry of large-scale structure formation in
real space by analytically integrating over orientations.
The linear input power spectrum can thereby have an arbitrary
functional
form as long as it can be represented on a high-resolution,
one-dimensional grid that is used for one-dimensional FFTs. In
particular, the method can easily resolve the imprint of baryonic acoustic
oscillations, BAO, on the initial power spectrum (see
Section~\ref{se:Pkshape}).
This is crucial for providing state-of-the-art model predictions for
the nonlinear evolution of BAO features in $\Lambda$CDM models and
extensions thereof.

Our goal in this paper is to generalize the \FFPT~approach introduced in \cite{MarcelZvonimirPat1603} to higher order in large-scale structure perturbation theory, specifically to the 2-loop power spectrum, corresponding to next-to-next-to-leading order in the linear mass density.  
This generalization is important to test the applicability of the fast \FFPT~framework of \cite{MarcelZvonimirPat1603} beyond 1-loop power spectrum integrals.
It should also help to make 2-loop perturbation theory more practically useable, for example to constrain cosmological parameters from a given dataset with only little computational cost.

While \FFPT~relies on exact analytical reformulations of the relevant 2-loop integrals, a viable alternative to reduce computational cost is to evaluate approximations of those integrals.
As demonstrated by Refs.~\cite{RegPT1208,Foreman1606},
 this can be achieved by Taylor expanding around a fiducial cosmological model, or by pre-computing integrals for a fiducial cosmology with high precision and then computing corrections for another cosmology with lower precision. 
The accuracy level and robustness of such approximate methods needs to be checked for every application, e.g.~when accounting for halo biasing, redshift space distortions or extensions of the basic $\Lambda$CDM model.

Although we share the same motivation and goals with Refs.~\cite{RegPT1208,Foreman1606}, our exact \FFPT~method is technically completely different and therefore complementary in practice, providing a useful path for cross-checks.
It would also be interesting to combine the ideas of \cite{RegPT1208,Foreman1606} and our method in the future,
particularly if the goal is to compute the 2-loop power spectrum robustly for different cosmological parameters at the sub-percent level precision that is needed to realize the full scientific potential of future LSS surveys.

For clarity we will focus on the standard 2-loop integrals for the matter power spectrum in SPT. 
However, our formalism can also handle halo bias, 
redshift space distortions (RSD), 
effects from the relative velocity between dark matter and baryons \cite{RelVelo1005}, or corrections from the effective field theory of large-scale structure \cite{2012JCAP...07..051B,2012JHEP...09..082C}, because the relevant integrals have the same form as the ones we consider here.
For example, halo bias can be included simply by modifying the perturbative $F_n$ kernels that enter the loop integrals (see Section~\ref{se:bias}),
while RSD effects amount to including additional velocity correlators
involving velocity kernels $G_n$
 (see Section~\ref{se:rsd}).
In principle it should also be possible to generalize the formalism to higher-order statistics beyond the power spectrum.
Our method should also work for cosmological models beyond $\Lambda$CDM as long as analytical expressions for perturbative kernels exist (see \cite{FasielloVlah1604} for recent progress in this direction). 
For models that do not allow for analytical perturbative kernels one instead has to resort to alternative approaches, for example computing kernels fully numerically.
While this is possible for subsets of 2-loop contributions by storing kernels on grids \cite{Taruya1606}, it is not clear if fourth or fifth order kernels could be included efficiently in such an approach.

Our paper is organized as follows. 
To get intuition, we first introduce higher-order corrections to 2-point statistics in a simple perturbative toy model in Section~\ref{se:background}.
In Section~\ref{se:P2LoopNoInvLaplace} we generalize this to a sub-class of simple 2-loop SPT power spectrum corrections that do not involve inverse Laplacians.
We then generalize this to account for a single inverse Laplacian  
in Section~\ref{se:P2LoopWITHInvLaplace}, and multiple inverse Laplacians
in Section~\ref{se:P2LoopManyInvLaplacians}.
In Section~\ref{se:Comments} we comment on the applicability of the method,
and extensions to e.g.~biased tracers.
Finally, we conclude in Section~\ref{se:conclusions}.
Appendices provide background material, derivations, and show how some of the general results simplify further
for the special case of scaling universes with power law
initial power spectrum.

\subsection*{Conventions and notation}
Throughout our paper, $\vk$ and $\vq$ refer to Fourier space, whereas $\vr$ and $\vx$ refer to position space.
We use the following shorthand notation for Fourier space integrals:
\begin{align}
  \label{eq:28}
  \int_{\vq} \equiv \int\frac{\d^3\vq}{(2\pi)^3}.
\end{align}
Hats denote unit vectors, e.g.~$\hat\vq=\vq/q$, where $q=|\vq|$.
$P_\lin$ denotes the linear matter density power spectrum, whereas $\mathsf{P}_\ell$ refers to Legendre polynomials.
We sometimes abbreviate indices of spherical harmonics as $\vl=(\ell,m)$ and use the shorthand notation $\sum_{\vl}^{l_\mathrm{max}}=\sum_{\ell=0}^{\ell_\mathrm{max}}\sum_{m=-\ell}^\ell$.
Spherical harmonics are normalized so that $\int\d\Omega_{\hat\vq}Y_{\ell m}(\hat\vq)Y^*_{\ell'm'}(\hat\vq)=\delta_{\ell\ell'}\delta_{mm'}$ and $Y_{00}(\hat\vq)=(4\pi)^{-1/2}$.
We highlight the most important results of our paper in boxed equations.

\section{Perturbative corrections to the 2-point correlation function}
\label{se:background}

In this section we introduce higher-order corrections to the matter 2-point correlation function in a simple toy model, which is useful to get intuition for the full corrections discussed later.

\subsection{Perturbative 2-point correlation function: Overview of terms in a toy model}
\label{se:toy2loop}

The approach of Eulerian standard perturbation theory (SPT) to solve the fluid equations for the large-scale dark matter overdensity is to expand this overdensity and the velocity perturbatively in the linear overdensity $\delta_1$,
\begin{align}
  \label{eq:16}
\delta(\vx)=\sum_{n=1}^\infty \delta_n(\vx).
\end{align}
Here, the $n$-th order contribution $\delta_n$ to the full nonlinear overdensity is of order $(\delta_1)^n$.
It has a known analytical form that follows from the fluid equations in an expanding universe.
The 2-point correlation function or power spectrum of the nonlinear density is then given by summing up contributions at different orders:
\begin{align}
  \la\delta\delta\ra \;=\; &
\underbrace{\la\delta_1\delta_1\ra }_{\text{tree-level}} 
+
\underbrace{2\la\delta_1\delta_3\ra 
  +\la\delta_2\delta_2\ra}_{\text{1-loop}}\non\\
&+
\underbrace{2\la\delta_1\delta_5\ra +
  2\la\delta_2\delta_4\ra +
  \la\delta_3\delta_3\ra}_\text{2-loop}
+
\underbrace{\mathcal{O}(\delta_1^8)}_\text{higher loops}.
\label{eq:correlTerms1}
\end{align}
The first term is the leading-order contribution, which is usually called the tree-level contribution because the corresponding Feynman diagram does not involve any loops. 
The next two terms are the 1-3 correlation between the linear and third order density,
and the 2-2 correlation between the two second order densities.
These are next-to-leading order contributions to the power spectrum. The Feynman diagrams of these 1-loop terms involve a single loop. 
The next three terms, corresponding to 1-5, 2-4 and 3-3 correlations, are next-to-next-to-leading-order terms, corresponding to Feynman diagrams with two loops that will be the focus of our paper.

These 2-loop integrals are typically studied in Fourier instead of position space. 
This has the advantage that differential operators like gradients or inverse Laplacians turn into analytical expressions of Fourier wavevectors, which are simple to write down and evaluate. 
However, working in Fourier space comes at the expense of introducing convolution integrals that would be simpler products of fields in position space. 

Since both the differential operators in position space and the convolutions in Fourier space represent substantial complications to typical calculations, we start with a simple but unphysical \emph{toy model} where we ignore all differential operators to simplify position space calculations.
Specifically, let us assume for a moment that the $n$-th order density is just the $n$-th power of the linear density,
\begin{align}
  \label{eq:ToyDefn}
  \delta_n(\vx)\equiv [\delta_1(\vx)]^n.
\end{align}
In this toy model, the 1-3 part of the 1-loop contribution to the 2-point correlation function is
\begin{align}
  \label{eq:18}
  \big\la\delta_1(\vx)\,\delta_3(\vx')\big\ra &\;=\; 
3\,
{
\contraction{\big\la}{\delta_1}{(\vx)\,}{\delta_1}
\contraction{\big\la\delta_1(\vx)\,\delta_1(\vx')}{\delta_1}{(\vx')}{\delta_1}
\big\la\delta_1(\vx)\,\delta_1(\vx')\delta_1(\vx')\delta_1(\vx')\big\ra
}\non\\
&\;=\; 
3\,\xi(r)\xi(0)
\end{align}
where the position $\vx'\equiv \vx+\vr$ is separated by a distance $r$ from $\vx$.
We also defined $\xi(r)=\la\delta_{1}(\vx)\delta_1(\vx')\ra$ as the 2-point correlation function of the linear density, with $\xi(0)$ representing the correlation at zero lag $r=0$.
Similarly, we obtain for the 2-2 contribution to the 2-point correlation function
\begin{align}
  \label{eq:18b}
  \big\la\delta_2(\vx)\,\delta_2(\vx')\big\ra 
&\,=\,
2
{
\contraction{\big\la}{\delta_1}{(\vx)\delta_1(\vx)\,}{\delta_1}
\bcontraction{\big\la\delta_1(\vx)}{\delta_1}{(\vx)\,\delta_1(\vx')}{\delta_1}
\big\la\delta_1(\vx)\delta_1(\vx)\,\delta_1(\vx')\delta_1(\vx')\big\ra  
}
\non\\&\quad\;
+
{
\contraction{\big\la}{\delta_1}{(\vx)}{\delta_1}
\contraction{\big\la\delta_1(\vx)\delta_1(\vx)\,}{\delta_1}{(\vx')}{\delta_1}
\big\la\delta_1(\vx)\delta_1(\vx)\,\delta_1(\vx')\delta_1(\vx')\big\ra
}
\non\\
&\,=\,
2\,[\xi(r)]^2+[\xi(0)]^2,
\end{align}

We can calculate similar expressions for 2-loop contributions in this simple toy model. 
The 1-5 contribution is given by a linear correlation function at nonzero separation $r$ multiplied by the square of the zero lag term $\xi(0)$,
\begin{align}
  \big\la \delta_1(\vx)\,\delta_5(\vx')\big\ra
&=
15\,
{
\contraction{\big\la}{\delta_1}{(\vx)\,}{\delta_1}
\contraction{\big\la\delta_1(\vx)\,\delta_1(\vx')}{\delta_1}{(\vx')}{\delta_1}
\contraction{\big\la\delta_1(\vx)\,\delta_1(\vx')\delta_1(\vx')\delta_1(\vx')}{\delta_1}{(\vx')}{\delta_1}
\big\la\delta_1(\vx)\,\delta_1(\vx')\delta_1(\vx')\delta_1(\vx')\delta_1(\vx')\delta_1(\vx')\big\ra
}\non\\
&=
15\,\xi(r)[\xi(0)]^2.
  \label{eq:xi15toy}
\end{align}
The 2-4 contribution has two qualitatively different contractions,
\begin{align}
  \big\la\delta_2(\vx)\,\delta_4(\vx')\big\ra
&=
12
{
\contraction{\big\la}{\delta_1}{(\vx)\delta_1(\vx)\,}{\delta_1}
\bcontraction{\big\la\delta_1(\vx)}{\delta_1}{(\vx)\,\delta_1(\vx')}{\delta_1}
\contraction{\big\la\delta_1(\vx)\delta_1(\vx)\,\delta_1(\vx')\delta_1(\vx')}{\delta_1}{(\vx')}{\delta_1}
\big\la\delta_1(\vx)\delta_1(\vx)\,\delta_1(\vx')\delta_1(\vx')\delta_1(\vx')\delta_1(\vx')\big\ra
}\non\\
&\quad
+
3
{
\contraction{\big\la}{\delta_1}{(\vx)}{\delta_1}
\contraction{\big\la\delta_1(\vx)\delta_1(\vx)\,}{\delta_1}{(\vx')}{\delta_1}
\contraction{\big\la\delta_1(\vx)\delta_1(\vx)\,\delta_1(\vx')\delta_1(\vx')}{\delta_1}{(\vx')}{\delta_1}
\big\la\delta_1(\vx)\delta_1(\vx)\,\delta_1(\vx')\delta_1(\vx')\delta_1(\vx')\delta_1(\vx')\big\ra
}
\non\\
\label{eq:xi24toy}
&=
12\,[\xi(r)]^2\xi(0)
+3\,[\xi(0)]^3.
\end{align}
Finally, the 3-3 contribution is
\begin{align}
  \big\la\delta_3(\vx)\,\delta_3(\vx')\big\ra
&=
6\,
{
\contraction[2.25ex]{\big\la}{\delta_1}{(\vx)\delta_1(\vx)\delta_1(\vx)\,\delta_1(\vx')\delta_1(\vx')}{\delta_1}
\contraction[1.5ex]{\big\la\delta_1(\vx)}{\delta_1}{(\vx)\delta_1(\vx)\,\delta_1(\vx')}{\delta_1}
\contraction[0.75ex]{\big\la\delta_1(\vx)\delta_1(\vx)}{\delta_1}{\,(\vx)}{\delta_1}
\big\la\delta_1(\vx)\delta_1(\vx)\delta_1(\vx)\,\delta_1(\vx')\delta_1(\vx')\delta_1(\vx')\big\ra
}
\non\\
&\quad +
9\,
{
\contraction{\big\la}{\delta_1}{(\vx)}{\delta_1}
\contraction{\big\la\delta_1(\vx)\delta_1(\vx)}{\delta_1}{(\vx)\,}{\delta_1}
\contraction{\big\la\delta_1(\vx)\delta_1(\vx)\delta_1(\vx)\,\delta_1(\vx')}{\delta_1}{(\vx')}{\delta_1}
\big\la\delta_1(\vx)\delta_1(\vx)\delta_1(\vx)\,\delta_1(\vx')\delta_1(\vx')\delta_1(\vx')\big\ra
}\non\\
\label{eq:xi33toy}
&=
6\,[\xi(r)]^3
+9\,\xi(r)[\xi(0)]^2.
\end{align}
In the toy model of \eqq{ToyDefn}, the 1- and 2-loop integrals thus only involve products of the 2-point correlation function $\xi(r)$ and the zero-lag correlation $\xi(0)$.
The computational cost of evaluating 1- and 2-loop integrals is therefore trivial in this toy model.  

It is not clear, however, if such a simple form of 1- and 2-loop integrals can also be obtained if we work with the full physical $n$-th order density perturbations that involve gradient and inverse Laplacian operators. 
While this has recently been shown to be the case for 1-loop integrals in \cite{MarcelZvonimirPat1603} (also see \cite{Mccollough1202,sherwinZaldarriaga,OSUloops1603}), it is not clear if 2-loop integrals allow similar simplifications. 
Addressing this question is the main goal of our paper.

Note that the constant $[\xi(0)]^2$ and $[\xi(0)]^3$ terms in 2-2 and 2-4 correlations are not present if we consider only the connected part of the correlation functions, $\la AB\ra_c=\la AB\ra-\la A\ra\la B\ra$, or if we enforce the density to have zero average at each order, $\delta_n=\delta_1^n-\la\delta_1^n\ra$.

\subsection{Eulerian fluid and equations of motion}

The toy model where the $n$-th order density perturbation is just the linear density raised to the $n$-th power is not physical because it does not solve the equations of motion of DM in an expanding background.
We briefly summarize here how to generalize the perturbative expansion so that it solves these equations (see \cite{bernardeauReview} for a review).

The relevant fluid equations can be written as the Fourier transform of the continuity equation,
\begin{align}
&\dot{\df}(\vec k, \tau) + \tf ( \vec k , \tau)\non\\
&= - \int_{\vec k_1 \vec k_2} 
(2\pi)^3 \df^D (\vec k - \vec k_1 -\vec k_2) 
\alpha (\vec k_1, \vec k_2)
\df(\vec k_1, \tau) \tf ( \vec k_2 , \tau),
\label{eq:ContEoM}
\end{align}
and the Fourier transform of the divergence of the Euler equation,
\begin{align}
&\dot{\tf}(\vec k, \tau) + \mathcal {H}(\tau) \tf ( \vec k , \tau) 
+ \frac{3}{2} \mathcal{H}(\tau)^2 \Omega_m(\tau) \df(\vec k, \tau)\non\\
&= - \int_{\vec k_1 \vec k_2} (2\pi)^3 \df^D (\vec k - \vec k_1 -\vec k_2) \beta (\vec k_1, \vec k_2)
\tf(\vec k_1, \tau) \tf ( \vec k_2 , \tau),
\label{eq:EulerEoM}
\end{align}
where $\delta$ is the matter overdensity, $\theta=\nabla\cdot\mathbf{v}$ is the velocity divergence, 
 $\dot{} = d / d \tau$ is the time derivative relative to conformal time, and $\mathcal{H}$ is the conformal Hubble parameter. 
We have also introduced the abbreviations
\begin{align}
 \alpha (\vec k_1, \vec k_2) &= \frac{ (\vec k_1 + \vec k_2) \cdot \vec k_2}{ k_2^2 }
 = 1 + \frac{1}{k_2^2}(\vec k_1 \cdot \vec k_2), \non\\
 \beta (\vec k_1, \vec k_2)  &= \frac{ (\vec k_1\cdot \vec k_2) |\vec k_1 + \vec k_2|^2 }{ 2 k_1^2 k_2^2 } \non\\
& = \frac{ \vec k_1\cdot \vec k_2}{ 2 k_1^2 k_2^2 } \lb k_1^2 + k_2^2 + 2 (\vec k_1 \cdot \vec k_2) \rb.
\end{align}
These kernels can be interpreted in position space by noting that multiplication with a wavevector $\vk$ corresponds to taking the gradient, whereas multiplication with $\vk/k^2$ corresponds to the gradient of the inverse Laplacian, i.e.~the gradient of a potential.

The equations of motion can be solved with the perturbative ansatz
\begin{align}
  \label{eq:7}
  \delta = \sum_{n=0}^\infty \delta_n,
\end{align}
where the $n$-th order perturbation in Fourier space is an $n$-fold convolution of the linear density $\delta_1$ filtered by a (symmetrized) kernel $F_n^{(s)}$,
\begin{align}
  \label{eq:13}
  \delta_n(\vk) = F_n^{(s)} \delta_1 * \cdots * \delta_1,
\end{align}
or writing this more explicitly,
\begin{align}
  \label{eq:17}
  \delta_n(\vk) = & \int_{\vq_1\cdots\vq_n} (2\pi)^3\delta_D(\vk-\vq_1\cdots-\vq_n) \non\\
& \quad \times F_n^{(s)}(\vq_1,\dots,\vq_n) \delta_1(\vq_1)\cdots\delta_1(\vq_{n}).
\end{align}
A similar expression follows for the velocity divergence $\theta$.

Explicit expressions for the $F_n^{(s)}$ kernels can be obtained from recursion relations that follow from the equations of motion; see Appendix~\ref{se:SPTreview}.
For our purposes, however, we only need to know the general form of the $F_n$ filter kernels.
This is determined by the operators appearing in the fluid equations of motion, involving e.g.~gradients and inverse Laplacians.
Indeed, the $n$-th order kernel $F_n^{(s)}(\vq_1,\dots,\vq_n)$ involves only sums of products of the following simple `building block' operators:
\begin{align}
 & F_n^{(s)}(\vq_1, \dots, \vq_n) \non\\
&\;\sim \;
\left\{
|\vq_i|^{n_i},
(\hat\vq_i\cdot\hat\vq_j)^{\ell_i},
\frac{1}{|s_1\vq_1+\cdots+s_n\vq_n|^2}
\right\},
\label{eq:FnBuildingBlocks}
\end{align}
where $n_i$ are integers, $l_i$ are non-negative integers, and
$s_i\in\{-1,0,1\}$. 
The last operator in \eqq{FnBuildingBlocks} corresponds to an inverse Laplacian.
The velocity kernels $G_n^{(s)}$ involve the same building blocks.

The simple toy model calculation from Section~\ref{se:toy2loop} thus needs to be refined by including these building blocks for the $n$-th order perturbation in Fourier space. 
Except for the inverse Laplacians, which require more work and will be discussed in a later section, this is relatively straightforward, as we will show next.

\section{2-loop power spectrum contributions without inverse Laplacians}
\label{se:P2LoopNoInvLaplace}

In this section we discuss contributions to the 2-loop matter power spectrum in Eulerian standard perturbation theory (SPT) that do not involve inverse Laplacians.  
Sections~\ref{se:P2LoopWITHInvLaplace} and \ref{se:P2LoopManyInvLaplacians} will generalize the results to account for such inverse Laplacians.

\begin{figure}[tb]
\centerline{
\includegraphics[width=0.4\textwidth]{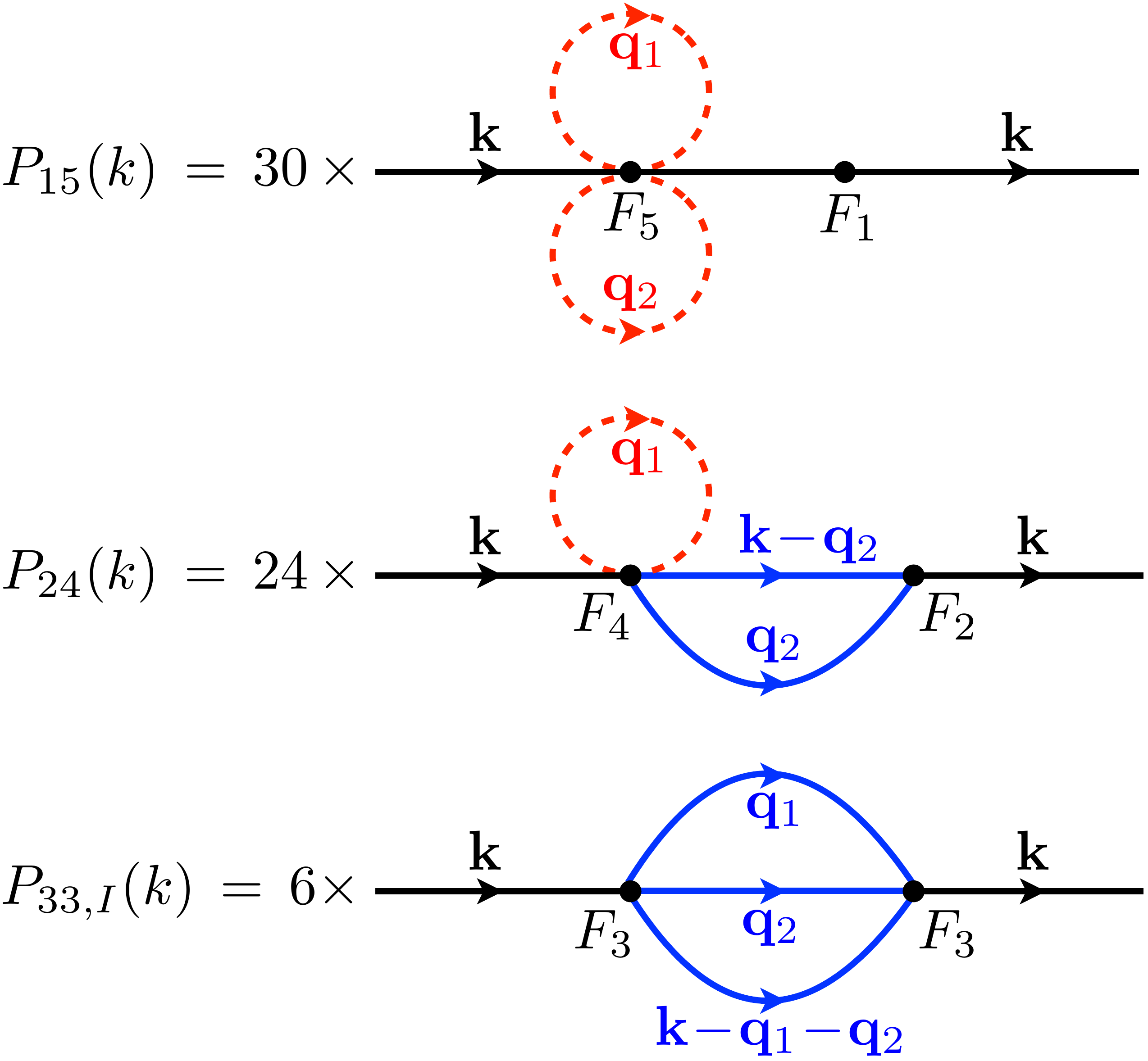}}
\caption{Diagrammatic representation of nontrivial 2-loop contributions to the dark matter power spectrum in standard Eulerian perturbation theory.
Tadpole subdiagrams (red dashed) are evaluated at the same point, leading to zero-lag correlations $\xi(0)$.
`Connector' subdiagrams (blue) are evaluated at two different points, leading to correlations $\xi(r)$ at non-zero separation $r$.
}
\label{fig:TwoLoopDiagrams}
\end{figure}

\begin{table*}[t]
\centering
 \begin{tabular}{lr}
\toprule
$(\ell_1,\ell_2,\ell_3)$ & $\mathcal{M}_{0}$\\\colrule
$(0,0,0)$&$1$\\
$(0,0,2)$&$1/3$\\
$(0,2,0)$&$1/3$\\
$(0,2,2)$&$1/9$\\
$(1,1,1)$&$1/9$\\
$(2,0,0)$&$1/3$\\
$(2,0,2)$&$1/9$\\
$(2,2,0)$&$1/9$\\
$(2,2,2)$&$11/225$\\
\botrule
\end{tabular}
$\qquad\qquad $
 \begin{tabular}{lcr}
\toprule
$(\ell_1,\ell_2,\ell_3)$ & $L$ & $\mathcal{M}_{1}$\\\colrule
$(0,0,0)$&$0$&$1$\\
$(0,1,1)$&$1$&$-1/3$\\
$(1,0,0)$&$1$&$-1$\\
$(1,1,1)$&$0$&$1/9$\\
$(1,1,1)$&$2$&$2/9$\\
\botrule\\
$ $\\
$ $\\
$ $\\
\end{tabular}
$\qquad\qquad $
 \begin{tabular}{lcr}
\toprule
$(\ell_1,\ell_2,\ell_3)$ & $(L,L')$ & $\mathcal{M}_{2}$\\\colrule
$(0,0,0)$&$(0,0)$&$1$\\
$(0,0,1)$&$(1,0)$&$-1$\\
$(0,1,1)$&$(0,1)$&$-1/3$\\
$(0,1,1)$&$(2,1)$&$-2/3$\\
$(1,1,1)$&$(0,0)$&$1/9$\\
$(1,1,1)$&$(0,2)$&$2/9$\\
$(1,1,1)$&$(2,0)$&$2/9$\\
$(1,1,1)$&$(2,2)$&$4/9$\\
\botrule\\
\end{tabular}
$\qquad\qquad $
 \begin{tabular}{lcr}
\toprule
$(\ell_1,\ell_2,\ell_3)$ & $(L_1,L_2,L_3)$ & $\mathcal{M}_{3}$\\\colrule
$(0,0,0)$&$(0,0,0)$&1\\
$(0,0,1)$&$(1,1,0)$&-1\\
$(0,1,1)$&$(0,1,1)$&-1/3\\
$(0,1,1)$&$(2,1,1)$&2/3\\
$(1,1,1)$&$(0,0,0)$&1/9\\
$(1,1,1)$&$(0,2,2)$&2/9\\
$(1,1,1)$&$(2,0,2)$&2/9\\
$(1,1,1)$&$(2,2,0)$&2/9\\
$(1,1,1)$&$(2,2,2)$&-2/9\\
\botrule
\end{tabular}
\caption{The angular structure of 2-loop integrands leads to coupling factors $\mathcal{M}_n$ between three angular momenta $(\ell_1,\ell_2,\ell_3)$ and $n$ angular momenta $L$.
The table shows all nonzero coupling factors $\mathcal{M}_{0}(\ell_1,\ell_2,\ell_3)$ for $\ell_i\le 2$,
$\mathcal{M}_{1}(\ell_1,\ell_2,\ell_3;L)$ for $\l_i\le 1$,
as well as 
 $\mathcal{M}_{2}(\ell_1,\ell_2,\ell_3;L,L')$ and $\mathcal{M}_{3}(\ell_1,\ell_2,\ell_3;L_1,L_2,L_3)$ for $\ell_1\le \ell_2\le \ell_3\le 1$.
See Appendix~\ref{se:CouplingFactors} for definitions of these factors.
}
\label{tab:M33}
\end{table*}

The form of the 2-loop corrections in full SPT is rather similar to the simple toy model from the last section.  
The final expressions therefore have a similar form to above, involving zero-lag terms $\xi(0)$ and correlation terms $\xi(r)$ at nonzero separation $r$. 
This can also be seen from the diagrammatic representation of the 2-loop integrals in Fig.~\ref{fig:TwoLoopDiagrams}, where `tadpole' subdiagrams (red dashed) lead to zero-lag terms $\xi(0)$ and `connector' subdiagrams (blue) lead to correlation terms $\xi(r)$.

\subsection{1-5 correlations}
We start with the 1-5 contribution to the power spectrum that arises from the correlation between the linear and fifth-order density. 
From the toy model result of \eqq{xi15toy} and the diagram in Fig.~\ref{fig:TwoLoopDiagrams} we expect this to be of the form $\xi(r)[\xi(0)]^2$, or $P_\mathrm{lin}(k)[\xi(0)]^2$ in Fourier space.
Explicitly, the 1-5 correlation in Fourier space is
\begin{align}
  \label{eq:24}
  P_{15}(k) \;=\;&
30\,
{
\contraction{\big\la}{\delta_1}{\, F^{(s)}_5\,}{\delta_1}
\contraction{\big\la\delta_1\, F^{(s)}_5\,\delta_1 *}{\delta_1}{*}{\delta_1}
\contraction{\big\la\delta_1\, F^{(s)}_5\,\delta_1 * \delta_1 * \delta_1 *}{\delta_1}{*}{\delta_1}
\big\la\delta_1\, F^{(s)}_5\,\delta_1 * \delta_1 * \delta_1 * \delta_1 * \delta_1 \big\ra}
\non\\
\;=\;&
30\, P_\lin(k)\int_{\vq_1\vq_2}F^{(s)}_5(\vk,\vq_1,-\vq_1,\vq_2,-\vq_2) 
\non\\
& \times P_\lin(q_1)P_\lin(q_2).
\end{align}
The prefactor arises from 15 possibilites to form the contraction multiplied by two because $\la\delta_1\delta_5\ra=\la\delta_5\delta_1\ra$.
The $F_5$ kernel consists of the building block operators listed in \eqq{FnBuildingBlocks}, and its angular structure can be parameterized by angular products between the arguments of the $F_5$ kernel, i.e.~the edges attached to the $F_5$ vertex in Fig.~\ref{fig:TwoLoopDiagrams}.
The most general form of such 1-5 contributions, ignoring inverse Laplacians for now, reduces to the following simple and fast-to-evaluate form (see endnote \footnote{The right hand side of \eqq{P15GeneralIntegralNoDenomi} follows by decomposing scalar products between wavevectors into spherical harmonics using \eqq{ScalProdInYlms}.  Then angular integrals over $\hat{\vq}_1$ and $\hat{\vq}_2$ follow from orthogonality of $Y_{lm}$'s, and $\sum_{m'}Y_{\ell' m'}(\hat\vk)Y^*_{\ell' m'}(\hat\vk)= (2\ell'+1)/(4\pi)$ 
gives \eqq{P15GeneralIntegralNoDenomi}.
})
\begin{empheq}[box=\fbox]{align}
&k^{n_0}P_\lin(k)
\int_{\vq_1\vq_2} (\hat\vq_1\cdot\hat\vq_2)^{\ell_0} (\hat\vk\cdot\hat\vq_1)^{\ell_1} (\hat\vk\cdot\hat\vq_2)^{\ell_2}\non\\
& 
\times q_1^{n_1}P_\lin(q_1)\,q_2^{n_2}P_\lin(q_2)\non\\
& \quad =\,
k^{n_0}P_\lin(k)  \,
\mathcal{M}_{0}(\ell_0,\ell_1,\ell_2)\, 
 \xi^{0}_{n_1}(0)\,\xi^{0}_{n_2}(0).
  \label{eq:P15GeneralIntegralNoDenomi}
\end{empheq}
As expected from the toy model \eqq{xi15toy}, the linear power spectrum $P_\lin(k)$ is multiplied by a $k$-independent product of two zero-lag correlations,
\begin{align}
  \label{eq:xin0}
  \xi_n^0(0) = \int_0^\infty\frac{\d q}{2\pi^2}\,q^{2+n} P_\lin(q).
\end{align}
These are fast to evaluate, either as a 1D integral over the linear power spectrum or by selecting the $r=0$ entry of a 1D Hankel transform with $l=0$.
The coupling factor $\mathcal{M}_{0}$ in \eqq{P15GeneralIntegralNoDenomi} is a number defined by \eqq{M15} in Appendix~\ref{se:CouplingFactors}.
The first few values are given in Table~\ref{tab:M33}.

\subsection{2-4 correlations}
We proceed with 2-4 correlations. 
From the toy model result of \eqq{xi24toy}, we expect them to be of the form $\xi(0)[\xi(r)]^2$ and $[\xi(0)]^3$, where only the former depends on the separation and contributes to the Fourier space power spectrum at nonzero wavenumber.
Explicitly, this 2-4 contribution to the power spectrum is
\begin{align}
  P_{24}(k) &=
24\,
{
\contraction{\big\la (F^{(s)}_4\,}{\delta_1}{*}{\delta_1}
\contraction{\big\la (F^{(s)}_4\,\delta_1*\delta_1*}{\delta_1}{*\delta_1)\, (F^{(s)}_2\,}{\delta_1}
\bcontraction{\big\la (F^{(s)}_4\,\delta_1*\delta_1*\delta_1*}{\delta_1}{)\, (F^{(s)}_2\,\delta_1*}{\delta_1}
\big\la (F^{(s)}_4\,\delta_1*\delta_1*\delta_1*\delta_1)\, (F^{(s)}_2\,\delta_1*\delta_1) \big\ra
}
\non\\
&=
24\int_{\vq_1\vq_2} 
F^{(s)}_4(\vq_1,-\vq_1,\vq_2,\vk\!-\!\vq_2)\,
F^{(s)}_2(\vq_2,\vk\!-\!\vq_2)\non\\
&\qquad \times P_\lin(q_1)\, P_\lin(q_2)\,P_\lin(|\vk\!-\!\vq_2|).
\end{align}
Introducing $\vq_3\equiv\vk\!-\!\vq_2$ with a Dirac delta, we obtain for 2-4 contributions without inverse Laplacians the following fast expression (see endnote \footnote{One way to derive the right-hand side of \eqq{P24generalNonDenomi} proceeds as follows:
Expand the Dirac delta in plane waves and each of those in spherical harmonics;
decompose scalar products between wavevectors in spherical harmonics; 
integrate over $\hat{\vq}_i$ and $\hat{\vr}$ using \eqq{GauntDef} and orthogonality of $Y_{lm}$'s;
sum over $m_1'$ and $m_2'$ using \eqq{ThreeJOrthoSumM1M2}; 
use \eqq{TwoYsInTermsOfOne} for $\sum_\VL \mathcal{G}_{LL_2L_2}^{MM_2,-M_2}Y^*_{LM}(\hat\vk)$; 
use Eqs.~\eq{Sum3JOverM} and \eq{3JSimpleCase}.
})
\begin{empheq}[box=\fbox]{align}
&\int_{\vq_1\vq_2\vq_3}
(2\pi)^3\delta_D(\vq_3\!-\!(\vk\!-\!\vq_2))
(\hat\vq_2\!\cdot\!\hat\vq_3)^{\ell_1} 
(\hat\vq_1\!\cdot\!\hat\vq_3)^{\ell_2}
\non\\
&
\times
(\hat{\vq}_1\!\cdot\!\hat{\vq}_2)^{\ell_3} 
q_1^{n_1}P_\lin(q_1)
q_2^{n_2} P_\lin(q_2)
q_3^{n_3}P_\lin(q_3)
\non\\
 &\;\;= \xi^0_{n_1}(0)\,(4\pi)^{3/2}
\int_0^\infty\d r\,r^2 j_0(kr)
\non\\
&\quad\;\;
\times\sum\limits_{L=0}^{\ell_1+\mathrm{min}(\ell_2,\ell_3)}
\mathcal{M}_{1}(\ell_1,\ell_2,\ell_3;L) \xi^{L}_{n_2}(r)\xi^{L}_{n_3}(r).
\label{eq:P24generalNonDenomi}
\end{empheq}
The right-hand side is similar to the $\xi(0)[\xi(r)]^2$ structure expected from the toy model result of \eqq{xi24toy} and the diagram in Fig.~\ref{fig:TwoLoopDiagrams}. 
The radial integral is the result of the angle-averaged 3D Fourier transform of $[\xi(r)]^2$; see \eqq{IntegrateExpOverAngle}.
This integral is weighted by a spherical Bessel function $j_\ell$ and is therefore a 1D Hankel transform.  
This can be evaluated efficiently and robustly with a 1D FFT using \textsf{FFTLog} \cite{hamiltonfftlog}.

In \eqq{P24generalNonDenomi} we defined a generalized correlation function
 $\xi^\ell_n$ as
\begin{eqnarray}
  \xi^\ell_n(r) 
  \label{eq:xidef}
&=& \int_0^\infty \frac{\d q}{2\pi^2}\,q^{2+n}\,j_l(qr)\,P_\lin(q).
\end{eqnarray}
This is related to the 2-point correlation between the linear density and a derivative or inverse Laplacian of the linear density \cite{MarcelZvonimirPat1603}.
Each $\xi^\ell_n(r)$ is a 1D Hankel transform of the linear power spectrum and can therefore again be computed with a 1D FFT.
The right-hand side of \eqq{P24generalNonDenomi} can therefore be evaluated using only 1D FFTs.
The coupling factors $\mathcal{M}_{1}$ are defined in \eqq{M24}, with some example values given in Table~\ref{tab:M33}.

\subsection{3-3 correlations}
The last contribution to the 2-loop power spectrum is the 3-3 correlation of two third-order densities.
From the toy model results of \eqq{xi33toy} we expect contributions of the forms $[\xi(r)]^3$ and $\xi(r)[\xi(0)]^2$.
It is well known that the latter term reduces to a term proportional to $[P_{13}(k)]^2/P_\mathrm{lin}(k)$, i.e.~it can be obtained directly from the $P_{13}$ 1-loop integral.
We therefore only consider the nontrivial 3-3 term which is of the form $[\xi(r)]^3$.
We expect that this part of the nontrivial 3-3 contribution to the power spectrum should become a 3D Fourier transform or 1D Hankel transform of $[\xi(r)]^3$.
Indeed, the nontrivial 3-3 power spectrum is
\begin{align}
  \label{eq:P33I}
&   P_{33,I}(k) \;=\;
6\,
{
\contraction[2.25ex]{\big\la (F^{(s)}_3\,}{\delta_1}{*\delta_1*\delta_1)(F^{(s)}_3\,\delta_1*\delta_1*}{\delta_1}
\contraction[1.5ex]{\big\la (F^{(s)}_3\,\delta_1*}{\delta_1}{*\delta_1)(F^{(s)}_3\,\delta_1*}{\delta_1}
\contraction[0.75ex]{\big\la (F^{(s)}_3\,\delta_1*\delta_1*}{\delta_1}{)(F^{(s)}_3\,}{\delta_1}
\big\la (F^{(s)}_3\,\delta_1*\delta_1*\delta_1)(F^{(s)}_3\,\delta_1*\delta_1*\delta_1)\big\ra
}\non\\
&\quad=\;
6\int_{\vq_1\vq_2\vq_3} (2\pi)^3\delta_D(\vk\!-\!\vq_1\!-\!\vq_2-\vq_3) \non\\
&\qquad\;\;\times
[F^{(s)}_3(\vq_1,\vq_2,\vq_3)]^2 P_\lin(q_1)P_\lin(q_2)P_\lin(q_3)
\end{align}
and contributions to this without inverse Laplacians reduce to (see endnote \footnote{
The right-hand side of \eqq{P33simple1} follows similarly to before: 
Expand the Dirac delta and scalar products between wavevectors in spherical harmonics, perform all angular integrations and sum over $m$'s.
One subtlety is that we averaged over orientations of the external $\vk$.
This does not affect the total physical power spectrum because we assume isotropy and ignored redshift space distortions (RSD). Even if individual contributions did depend on $\hat{\vk}$ we could still integrate over $\hat{\vk}$, because if $P=P_1+P_2$ then $4\pi P = \int_{\hat\vk} P=\int_{\hat{\vk}} P_1+\int_{\hat{\vk}} P_2$.
We note that RSD can be included in this framework by expressing the observable redshift space power spectrum in terms of isotropic statistics of the mass and velocity densities as discussed in Section~\ref{se:rsd}.
})
\begin{empheq}[box=\fbox]{align}
&\int\frac{\d\Omega_{\hat\vk}}{4\pi}
\int_{\vq_1\vq_2\vq_3} (2\pi)^3\delta_D(\vk\!-\!\vq_1\!-\!\vq_2\!-\!\vq_3)
\non\\
&\quad\times
(\hat\vq_2\!\cdot\!\hat\vq_3)^{\ell_1}
(\hat\vq_1\!\cdot\!\hat\vq_3)^{\ell_2}
(\hat\vq_1\!\cdot\!\hat\vq_2)^{\ell_3}
\prod_{i=1}^3 q_i^{n_i}P_\lin(q_i)
\non\\
& =
4\pi
\int_0^\infty\d r\,r^2 j_0(kr) 
\bigg[
\sum_{L_1=0}^{\ell_2+\ell_3}
\sum_{L_2=0}^{\ell_1+\ell_3}
\sum_{L_3=0}^{\ell_1+\ell_2}\non\\
&\quad
\mathcal{M}_{3}(\ell_1,\ell_2,\ell_3;L_1,L_2,L_3)
\xi^{L_1}_{n_1}(r)\,\xi^{L_2}_{n_2}(r)\,\xi^{L_3}_{n_3}(r)\bigg]
\label{eq:P33simple1}
\end{empheq}
As expected, this is a 1D Hankel transform of a finite sum of triple products of linear correlation functions $\xi^\ell_n(r)$, which can be computed using only 1D FFTs.

The right-hand side of \eqq{P33simple1} involves the coupling factor $\mathcal{M}_{3}$ defined by \eqq{M33def}. 
For $\ell_1\le
\ell_2\le \ell_3\le 1$ the only nonzero couplings to $L_i$ are listed
in Table \ref{tab:M33}.
For example, for
$(\ell_1,\ell_2,\ell_3)=(0,1,1)$ only two couplings are nonzero,
$(L_1,L_2,L_3)=(0,1,1)$ and $(2,1,1)$, so we only need to compute
$\xi^{0}_{n_1}$, $\xi^{2}_{n_1}$, $\xi^{1}_{n_2}$ and $\xi^{1}_{n_3}$
and one additional 1D Hankel transform to go back to Fourier space,
requiring five 1D FFTs in total.

\section{Single inverse Laplacian in the 2-loop power spectrum}
\label{se:P2LoopWITHInvLaplace}

For simplicity, we have ignored inverse Laplacian operators so far. 
In fact, however, they do appear in the perturbative solutions for the DM fluid because the continuity equation \eq{ContEoM} and the Euler equation \eq{EulerEoM} involve the gradient of the velocity divergence potential, $\nabla_i\nabla^{-2}\theta$.
When solving the equations perturbatively, we therefore encounter expressions like the inverse Laplacian of composite quadratic fields, e.g.~$\nabla^{-2}\theta^{(2)}\sim \nabla^{-2}[\delta_1^2]$.
In Fourier space, this is represented by terms like $|\vq_1+\vq_2|^{-2}$ (this can also be seen from the recursion relations for the perturbative Fourier space $F_n$ kernels; see Appendix~\ref{se:SPTreview}).
Such Fourier space factors can render the integrand of loop integrals nonseparable in the integration variables $\vq_1$ and $\vq_2$, so that the integrand cannot be written as a function of $\vq_1$ multiplied by a function of $\vq_2$.
This may seem problematic for the approach used in the previous section, because the 2-loop integrals do not straightforwardly split into an integral over $\vq_1$ multiplied by an integral over $\vq_2$.
In this section we show that it is still possible to
reduce 2-loop integrals with a single inverse Laplacian to 1D Hankel transforms that allow fast evaluation.
The more complicated case involving multiple inverse Laplacians will be discussed in Section~\ref{se:P2LoopManyInvLaplacians}.

\subsection{1-5 correlations with inverse Laplacians: Products of two correlation functions}

\subsubsection{Simple example}

We first generalize the 1-5 correlations from Eqs.~\eq{xi15toy} and \eq{P15GeneralIntegralNoDenomi} to the case with a nontrivial inverse Laplacian.
To see how the inverse Laplacian can look like in Fourier space, consider for example
\begin{align}
&
  \int\d^3\vr\,e^{i\vk\cdot\vr}
\big\la
{
\contraction{}{\delta_1}{(\vx)\,}{\delta_1}
\contraction{\delta_1(\vx)\,\delta_1(\vx')}{\delta_1}{(\vx')\delta_1(\vx')\nabla^{-2}\big[}{\delta_1}
\bcontraction{\delta_1(\vx)\,\delta_1(\vx')\delta_1(\vx')}{\delta_1}{(\vx')\nabla^{-2}\big[\delta_1(\vx')}{\delta_1}
\delta_1(\vx)\,\delta_1(\vx')\delta_1(\vx')\delta_1(\vx')\nabla^{-2}\big[\delta_1(\vx')\delta_1(\vx')\big]
}
\big\ra\non\\
&\quad =\,
-P_\lin(k)\int_{\vq_1\vq_2}\frac{1}{|\vq_1+\vq_2|^2}P_\lin(q_1)P_\lin(q_2),
\label{eq:P15InvLaplaceDemo}
\end{align}
where the right-hand side follows from 
\begin{align}
  \label{eq:ddcontraction}
\big\la
{\contraction{}{\delta_1}{(\vx)}{\delta_1}
 \delta_1(\vx)\delta_1(\vx')} 
\big\ra
= \int_\vq e^{-i\vq\cdot(\vx-\vx')}P_\lin(q).  
\end{align}
The particular inverse Laplacian in \eqq{P15InvLaplaceDemo} thus turns into $-|\vq_1+\vq_2|^{-2}$ in Fourier space.
To evaluate the resulting 2-loop integral over $\vq_1$ and $\vq_2$ efficiently, we introduce $\vq_3=\vq_1+\vq_2$ with a Dirac delta, integrate out all orientations and use \eqq{FTofInvLaplacian} to obtain
\begin{align}
  \label{eq:P15InvLaplSimple}
  \int_{\vq_1\vq_2}\frac{P_\lin(q_1)P_\lin(q_2)}{|\vq_1+\vq_2|^2} 
= \int_0^\infty \d r\,r\, \xi^0_0(r)\, \xi^0_0(r).
\end{align}
This is just a radial integral over the product of two correlation functions, which can be evaluated very efficiently.

\subsubsection{General case}

The most general form of the inverse Laplacian can be deduced from the arguments of the $F_5$ kernel in \eqq{P15GeneralIntegralNoDenomi} and the form of the kernel recursion relations in Appendix~\ref{se:SPTreview}; it is given by $|s_0\vk+s_1\vq_1+s_2\vq_2|^{-2}$, with parameters $s_i\in\{-1,0,1\}$ that parameterize on which fields the inverse Laplacian acts. 
Allowing also for nontrivial angular dependence in the integrand that arises from gradient operators $\nabla_i$, we obtain for the most general 1-5 contribution to the power spectrum 
(see endnote \footnote{To simplify the integral in \eqq{P15GeneralFast} we introduce the auxiliary variable $\vq_3\equiv s_0\vk+s_1\vq_1+s_2\vq_2$ using a Dirac delta. 
The right-hand side then follows by decomposing scalar products into spherical harmonics, and expanding the Dirac delta in plane waves and and those in spherical harmonics.
Performing angular integrations over $\hat{\vq}_i$, $\hat\vr$ and $\hat\vk$ using \eqq{GauntDef} leads to a product of four Gaunt integrals, whose sum over $m$'s is a Wigner 6-j symbol, giving \eqq{P15GeneralFast}.
We also used \eqq{FTofInvLaplacian}.
} for a derivation):
\begin{empheq}[box=\fbox]{align}
&k^{n_0} P_\lin(k)
\int\frac{\d\Omega_{\hat\vk}}{4\pi}
\int_{\vq_1\vq_2}
(\hat\vq_1\cdot\hat\vq_2)^{\ell_0} \,
(\hat\vk\cdot\hat\vq_1)^{\ell_1}
 \non\\
&\quad\times
 (\hat\vk\cdot\hat\vq_2)^{\ell_2}\,
\frac{q_1^{n_1}P_\lin(q_1)\,q_2^{n_2}P_\lin(q_2)}{|s_0\vk+s_1\vq_1+s_2\vq_2|^{2}}\non\\
& =
k^{n_0} P_\lin(k)
\sum_{L_0=0}^{\ell_1+\ell_2}
(\sgn\,s_0)^{L_0}
 \int_0^\infty\d r\,r\, j_{L_0}(|s_0|kr)
\non\\
&\quad\times
\Bigg[
\sum_{L_1=0}^{\ell_0+\ell_1}\sum_{L_2=0}^{\ell_0+\ell_2}
\mathcal{M}_{3}(\ell_0,\ell_1,\ell_2;L_0,L_2,L_1)
\non\\
&\qquad\quad\times
\xi^{L_1}_{n_1,s_1}(r)\,\xi^{L_2}_{n_2,s_2}(r)
\Bigg].
\label{eq:P15GeneralFast}
\end{empheq}

The right-hand side of \eqq{P15GeneralFast} is given by 1D Hankel transforms of products of two correlation functions $\xi$, which are themselves given by 1D Hankel transforms of the linear power spectrum.
Thus, using Eqs.~\eq{P15GeneralIntegralNoDenomi} and \eq{P15GeneralFast}, the calculation of the full $P_{15}(k)$ contribution to the 2-loop power spectrum at all $k$ can be obtained by a sequence of 1D Hankel transforms, which are fast to compute with 1D FFTs using \textsf{FFTLog} \cite{hamiltonfftlog}.

Note that \eqq{P15GeneralFast} is only meant to be applied for cases with nontrivial inverse Laplacians where at least two of $s_0,s_1,s_2$ are nonzero, because otherwise there is no nonseparable denominator and \eqq{P15GeneralIntegralNoDenomi} can be applied instead.
The right-hand side of \eqq{P15GeneralFast} involves the coupling factor $\mathcal{M}_{3}$  defined in \eqq{M33def} and listed in Table~\ref{tab:M33}.
It also involves the generalized correlation functions
\begin{align}
  \label{eq:xidefWithParam}
    \xi^\ell_{n,s}(r) \equiv (\sgn\,s)^l \int_0^\infty \frac{\d q}{2\pi^2}\,
q^{2+n}\,P_\lin(q)\,j_l(|s|qr),
\end{align}
where $s\in\{-1,0,1\}$. 
They reduce to the usual correlations $\xi^\ell_n(r)$ for $|s|=1$ and to zero-lag terms $\xi^0_n(0)$ for $s=0$, because $j_l(0)=\delta_{l0}$.
In the special case $\ell_i=n_i=s_0=0$ and $s_1=s_2=1$, \eqq{P15GeneralFast} reduces to the simple result of \eqq{P15InvLaplSimple}.

\subsection{2-4 correlations with inverse Laplacians} 

\subsubsection{Simple example}

\label{se:P24TrivialAngles}

We now turn to  2-4 correlations including inverse Laplacians.
Our main idea to evaluate these 2-loop integrals is to split them into 
nested 1-loop integrals that are much simpler to evaluate. 
For clarity we introduce this approach first for a simple special case in this section, discussing the fully general case in the subsequent section and in Appendix~\ref{app:P24WithInvLaplAndAnglesAppdx}.

The special case we consider is given by the first contraction of \eqq{xi24toy} if we include an inverse Laplacian acting on the squared linear density as
\begin{align}
&\int\d^3\vr\,e^{i\vk\cdot\vr}\,
  {
\contraction{\big\la}{\delta_1}{(\vx)\delta_1(\vx)\,}{\delta_1}
\contraction{\big\la\delta_1(\vx)\delta_1(\vx)\,\delta_1(\vx')}{\delta_1}{(\vx')\nabla^{-2}\big[}{\delta_1}
\bcontraction{\big\la\delta_1(\vx)}{\delta_1}{(\vx)\,\delta_1(\vx')\delta_1(\vx')\nabla^{-2}\big[\delta_1(\vx')}{\delta_1}
\big\la\delta_1(\vx)\delta_1(\vx)\,\delta_1(\vx')\delta_1(\vx')\nabla^{-2}\big[\delta_1(\vx')\delta_1(\vx')\big]\big\ra
  }
\non\\
& \quad =
-\int_{\vq_1\vq_2} 
\, \frac{P_\lin(q_1)\,P_\lin(q_2)\,
P_\lin(|\vk\!-\!\vq_2|)}{|\vq_1 + \vq_2|^{2}},
  \label{eq:P24generalSimple1a}
\end{align}
where the right-hand side follows from \eqq{ddcontraction}.
To speed up evaluation, the main idea is now to write the 2-loop integral \eq{P24generalSimple1a} as an outer $\vq_2$-integral over an inner tadpole integral over $\vq_1$:
\begin{align}
&\int_{\vq_1\vq_2} 
\, \frac{P_\lin(q_1)\,P_\lin(q_2)\,
P_\lin(|\vk\!-\!\vq_2|)}{|\vq_1 + \vq_2|^{2}}
\non\\
&\quad = \int_{\vq_2}P_\lin(q_2)
\underbrace{\left[\int_{\vq_1}\frac{P_\lin(q_1)}{|\vq_1+\vq_2|^2}\right]}_{P_\mathrm{tadpole}(q_2)}
P_\lin(|\vk\!-\!\vq_2|).
  \label{eq:P24generalSimple1b}
\end{align}
This reduces the 2-loop integral to two nested 1-loop integrals that are easy to evaluate.
In the diagrammatic representation of Fig.~\ref{fig:TwoLoopDiagrams}, this corresponds to evaluating the red tadpole subdiagram first, and then using the result to compute the blue subdiagram connecting the $F_4$ and $F_2$ vertices.

To see more specifically how \eqq{P24generalSimple1b} simplifies numerical evaluation, we write the inner tadpole integral as (see endnote \footnote{This follows by introducing $\vq_3=\vq_1+\vq_2$ with a Dirac delta, expanding this in spherical harmonics, performing all angular integrals, and using \eqq{FTofInvLaplacian}.})
\begin{align}
\label{eq:PtadpoleNoAngle}
  P_\mathrm{tadpole}(q_2) \;=\;
\,\int_0^\infty\d r\,r\,j_0(q_2r)\,\xi^0_0(r).
\end{align}
Then, we can evaluate the Fourier space convolution over $\vq_2$ in \eqq{P24generalSimple1b} as a product in position space, obtaining
\begin{align}
&\int_{\vq_1\vq_2} 
\, \frac{P_\lin(q_1)\,P_\lin(q_2)\,
P_\lin(|\vk\!-\!\vq_2|)}{|\vq_1 + \vq_2|^{2}}\non\\
&\quad = 
4\pi\int_0^\infty\d r\,r^2\, j_0(kr)\,\xi^0_0(r)\,\mathcal{T}(r).
  \label{eq:P24FastNoAngleDep}
\end{align}
This is a 1D Hankel transform of the product between the linear correlation function $\xi^0_0(r)$ and the 4-point-like correlation $\mathcal{T}(r)$.
The latter is defined as a 1D Hankel transform of the product of the linear power spectrum $P_\lin$ and the tadpole integral $P_\mathrm{tadpole}$:
\begin{align}
\label{eq:TauSimple}
  \mathcal{T}(r) \equiv \int_0^\infty\frac{\d q_2}{2\pi^2}
\,q_2^2\, j_0(q_2r)\, P_\lin(q_2)\,P_\mathrm{tadpole}(q_2).
\end{align}

Using \eqq{P24FastNoAngleDep}, the 2-loop integral of \eqq{P24generalSimple1a} can be computed from a given linear power spectrum with four 1D Hankel transforms in total, which is extremely fast.
Similar reductions of 2-loop integrals to two nested 1-loop integrals are also used in other contexts to simplify their evaluation; see e.g.~\cite{PeskinSchroederBook} for examples in quantum field theory.

\subsubsection{General case}
\label{se:P24WithInvLaplAndAngles}

The 2-loop integral \eq{P24generalSimple1a} from the last section is a special case in the sense that the integrand does not contain nontrivial angular dependence from terms like e.g.~$\hat\vq_1\cdot\hat\vq_2$. 
One of the main results of our paper is that the \FFPT~approach still works if such nontrivial angular dependence is included in the integrand. 
While this leads to additional coupling factors, the general strategy is the same as in the last section, i.e.~we split the 2-loop integral into two nested 1-loop integrals that can be evaluated as 1D Hankel transforms.
This is discussed in detail in Appendix~\ref{app:P24WithInvLaplAndAnglesAppdx}.
The final results for the nontrivial 2-4 correlations, given by Eqs.~\eq{P24GeneralFastCase1} and \eq{P24GeneralFastCase3}, involve only 1D Hankel transforms, which can be evaluated efficiently with a finite number of 1D FFTs using \textsf{FFTLog} \cite{hamiltonfftlog}.

\subsection{3-3 correlations with inverse Laplacians: $\xi$ times transformed $\xi^2$}
\label{se:P33withdenomi}

\subsubsection{Simple example}

We finally turn to the last remaining contribution to the 2-loop power spectrum arising from nontrivial 3-3 correlations.
These involve for example an inverse Laplacian acting on two linear densities in \eqq{xi33toy} as follows:
\begin{align}
&  \int\d^3 \vr\, e^{i\vk\cdot\vr}\,
{
\contraction[2.25ex]{\big\la}{\delta_1}{(\vx)\delta_1(\vx)\delta_1(\vx)\,\delta_1(\vx')\nabla^{-2}\big[\delta_1(\vx')}{\delta_1}
\contraction[1.5ex]{\big\la\delta_1(\vx)}{\delta_1}{(\vx)\delta_1(\vx)\,\delta_1(\vx')\nabla^{-2}\big[}{\delta_1}
\contraction[0.75ex]{\big\la\delta_1(\vx)\delta_1(\vx)}{\delta_1}{(\vx)\,}{\delta_1}
\big\la\delta_1(\vx)\delta_1(\vx)\delta_1(\vx)\,\delta_1(\vx')\nabla^{-2}\big[\delta_1(\vx')\delta_1(\vx')\big]\big\ra
}\non\\
&\quad =
-\int_{\vq_1\vq_2} \frac{P_\lin(q_1)P_\lin(q_2)P_\lin(|\vk\!-\!\vq_1\!-\!\vq_2|)}{|\vq_1+\vq_2|^2}.
  \label{eq:P33InvLaplaceContractions}
\end{align}
Again, the inverse Laplacian turns into $-|\vq_1+\vq_2|^{-2}$ in Fourier space. 
This 2-loop integral can be simplified to (see Appendix~\ref{se:P33SimpleDerivAppdx})
\begin{align}
& \int\frac{\d\Omega_{\hat\vk}}{4\pi}\int_{\vq_1\vq_2} \frac{P_\lin(q_1)P_\lin(q_2)P_\lin(|\vk\!-\!\vq_1\!-\!\vq_2|)}{|\vq_1+\vq_2|^2}
\non\\
&\quad\quad =
(4\pi)^4
\int_0^\infty\d r\,r^2\,j_0(kr)\,\xi^0_0(r)\,\mathbb{T}(r).
\label{eq:P33SimpleInvDenomi}
\end{align}
This is the Hankel transform of the product of the linear correlation function $\xi^0_0(r)$ and the 4-point like quantity $\mathbb{T}(r)$ defined by
\begin{align}
  \label{eq:bbTSimple}
  \mathbb{T}(r)  
& = \int_0^\infty\frac{\d q}{2\pi^2}\,j_0(qr)
\int_0^\infty\d r''(r'')^2j_0(qr'')\left[\xi^0_0(r'')\right]^2.
\end{align}
The latter is obtained by squaring the linear correlation function in position space, transforming the result to Fourier space using a Hankel transform, dividing by $q^2$, and transforming back to position space with another Hankel transform.
Therefore, the 2-loop integral of \eqq{P33SimpleInvDenomi} is essentially given by $\xi$ times a transform of $\xi^2$.
This is extremely numerically efficient.

\subsubsection{General case}

The most general form of integrals contributing to nontrivial 3-3 correlations follows by also including scalar products in the integrand, and introducing $\vk-\vq_1-\vq_2=\vq_3$ (see endnote \footnote{The most general inverse Laplacian operator allowed by the arguments of the $F_3$ kernel in the 3-3 correlation \eq{P33I} is $|s_1\vq_1+s_2\vq_2+s_3\vq_3|^{-2}$ with $s_i\in\{0,1\}$.
We can restrict ourselves to a single denominator of this form by using partical fraction decompositions if needed.
The only case not already covered by \eqq{P33simple1} arises if two $s_i$ are $1$ and the other one is $0$.
Without loss of generality, we consider $(s_1,s_2,s_3)=(1,1,0)$, corresponding to $|\vq_1+\vq_2|^{-2}$, or, slightly more generally, $|\vq_1+\vq_2|^{-n_4}$.
} and Appendix~\ref{se:P33DerivationAppendix}):
\begin{empheq}[box=\fbox]{align}
&\int\frac{\d\Omega_{\hat\vk}}{4\pi}\int_{\vq_1\vq_2\vq_3}(2\pi)^3\delta_D(\vk\!-\!\vq_1\!-\!\vq_2-\vq_3)
\non\\
&\, 
\times
(\hat\vq_2\cdot\hat\vq_3)^{\ell_1}
(\hat\vq_1\cdot\hat\vq_3)^{\ell_2}
(\hat\vq_1\cdot\hat\vq_2)^{\ell_3}
\frac{1}{|\vq_1+\vq_2|^{n_4}}\,
\non\\
& \,\times 
q_1^{n_1}P_\lin(q_1)
q_2^{n_2}P_\lin(q_2)
q_3^{n_3}P_\lin(q_3)
\non\\
&\quad=
(4\pi)^4
\int_0^\infty \d r\,r^2 j_0(kr)
\sum_{L_3=|\ell_1-\ell_2|}^{\ell_1+\ell_2} 
\non\\
&\qquad\quad
\xi^{L_3}_{n_3}(r)
\, \mathbb{T}_{L_3}^{\ell_1\ell_2\ell_3,n_1n_2n_4}(r).
  \label{eq:P33withDenomiFast}
\end{empheq}
The right-hand side is a 1D Hankel transform of a sum of products between linear correlation functions $\xi(r)$ and the 4-point-like quantity $\mathbb{T}(r)$. The latter is defined by applying two subsequent 1D Hankel transforms to a product of two correlation functions $\xi(r)$:
\begin{align}
& 
 \mathbb{T}_{L_3}^{\ell_1\ell_2\ell_3,n_1n_2n_4}(r)
\non\\
&
\equiv  \int_0^\infty\frac{\d q}{2\pi^2}\,q^{2-n_4} j_{L_3}(qr) 
\Bigg\{\int_0^\infty\d r''(r'')^2 j_{L_3}(qr'')
\non\\
&
\quad\times\Bigg[\sum_{L_1=|\ell_2-\ell_3|}^{\ell_2+\ell_3}\sum_{L_2=|\ell_1-\ell_3|}^{\ell_1+\ell_3}\mathcal{M}_{3}(\ell_1,\ell_2,\ell_3;L_1,L_2,L_3)\non\\
&\qquad\quad\times\xi^{L_1}_{n_1}(r'')\xi^{L_2}_{n_2}(r'')
\Bigg]\Bigg\},
\label{eq:XdefP33}
\end{align}
where the coupling factor $\mathcal{M}_{3}$ from \eqq{M33def} restricts the sums to be finite.

The general 3-3 correlation of \eqq{P33withDenomiFast} can thus be evaluated with a finite number of 1D Hankel transforms.
The structure is similar to the $[\xi(r)]^3$ structure obtained for 3-3 correlations without inverse Laplacians in Eqs.~\eq{xi33toy} and \eq{P33simple1}, but the outer-most integral in \eqq{XdefP33} effectively applies an inverse Laplacian to the product of two correlation functions $\xi(r)$ as expected from the contractions in \eqq{P33InvLaplaceContractions} and the simple example of \eqq{P33SimpleInvDenomi}.

\section{Multiple inverse Laplacians}
\label{se:P2LoopManyInvLaplacians}

Unfortunately, the full 2-loop power spectrum also involves contributions that have multiple nontrivial inverse Laplacians, corresponding to multiple nonseparable denominators in 2-loop integrands.
Since they involve 3d wavevectors one cannot simply separate the
denominators using a partial fraction decomposition and then apply the
machinery laid out in the last sections. 
Instead, we follow a somewhat different approach than in the rest of the paper. 
This reduces contributions with multiple nontrivial inverse Laplacians
to low-dimensional radial integrals.  
We explicitly show this for the case of trivial numerators in the
Fourier space integrals and indicate how more complicated numerators could in principle be generated from this. 

\subsection{1-5 correlations}

Based on explicit calculation of the $F_5$ kernel
that enters 1-5 contributions to the power spectrum, we consider
integrals of the general form
\begin{align}
  I_{15}(k,\valpha,\vbeta) =& \int_{\vq_1\vq_2}\frac{e^{i\valpha\cdot\vq_1}e^{i\vbeta\cdot\vq_2}}{q_1^{2n_1} |\vk+\vq_1|^{2n_1'}q_2^{2n_2}|\vk+\vq_2|^{2n_2'}}\non\\
\label{eq:I15def}
&\times  \frac{P_\lin(q_1)P_\lin(q_2)P_\lin(k)}{|\vq_1+\vq_2|^{2n_3}  |\vk+\vq_1+\vq_2|^{2n_3'}},
\end{align}
where $n_i,n_i'\leq 2$ and we introduced parameters $\valpha$ and $\vbeta$.
Nontrivial numerators can be generated by computing $I_{15}$ and taking
appropriate derivatives with respect to $\valpha$ and $\vbeta$ evaluated at
zero, although we do not explicitly do this here. 
Introducing a helper variable for 
$\vk+\vq_1+\vq_2$ with a Dirac delta, \eqq{I15def} reduces to
\begin{align}
  \label{eq:I15WithXiN}
  I_{15}(k,\valpha,\vbeta) =\,& P_\lin(k)
\int\d^3\vr\,e^{i\vk\cdot\vr}
\,\Xi^{(0)}_{n_3'n_3}(\vr,\vk)
\non\\
&\times
\Xi^{(1)}_{n_1n_1'}(\vr+\valpha,\vk)
\, \Xi^{(1)}_{n_2n_2'}(\vr+\vbeta,\vk).
\end{align}
The computation is therefore reduced to calculating
\begin{align}
  \label{eq:Xi0Def}
  \Xi^{(0)}_{nn'}(\vr,\vk) & \equiv  \int_\vq\,\frac{e^{i\vq\cdot\vr}}{q^{2n}|\vk+\vq|^{2n'}},\\
\label{eq:Xi1Def}
  \Xi^{(1)}_{nn'}(\vr,\vk) &\equiv \int_\vq\,\frac{e^{i\vq\cdot\vr}}{q^{2n}|\vk+\vq|^{2n'}}\,P_\lin(q),
\end{align}
which we will discuss below.

\subsection{2-4 correlations}

Similarly, 2-4 correlations with multiple
inverse Laplacians of the form
\begin{align}
  I_{24}(k,\valpha,\vbeta) =&
 \int_{\vq_1\vq_2}\frac{e^{i\valpha\cdot\vq_1}e^{i\vbeta\cdot\vq_2}}{q_1^{2n_1} |\vk+\vq_1|^{2n_1'}q_2^{2n_2}|\vk+\vq_2|^{2n_2'}}\non\\
\label{eq:I24def}
&\times  \frac{P_\lin(q_1)P_\lin(q_2)P_\lin(|\vk+\vq_2|)}{|\vq_1+\vq_2|^{2n_3}  |\vk+\vq_1+\vq_2|^{2n_3'}}
\end{align}
can be reduced to
\begin{align}
  \label{eq:I24WithXiN}
  I_{24}(k,\valpha,\vbeta) =&
 \int\d^3\vr\,e^{i\vk\cdot\vr}
\,\Xi^{(0)}_{n_3'n_3}(\vr,\vk)
\,\Xi^{(1)}_{n_1n_1'}(\vr+\valpha,\vk)\non\\
&\times \Xi^{(2)}_{n_2n_2'}(\vr+\vbeta,\vk).
\end{align}
Here we defined
\begin{align}
\label{eq:Xi2Def}
  \Xi^{(2)}_{nn'}(\vr,\vk) & \equiv  \int_\vq\,\frac{e^{i\vq\cdot\vr}}{q^{2n}|\vk+\vq|^{2n'}}\,P_\lin(q)P_\lin(|\vk+\vq|).
\end{align}

\subsection{3-3 correlations}

For 3-3 contributions to the power spectrum with multiple inverse Laplacians
we consider the general integral
\begin{align}
  I_{33}(k,\valpha,\vbeta) =& \int_{\vq_1\vq_2}\frac{e^{i\valpha\cdot\vq_1}e^{i\vbeta\cdot\vq_2}}{q_1^{2n_1} |\vk+\vq_1|^{2n_1'}q_2^{2n_2}|\vk+\vq_2|^{2n_2'}}\non\\
\label{eq:I33def}
&\times  \frac{P_\lin(q_1)P_\lin(q_2)P_\lin(|\vk+\vq_1+\vq_2|)}{|\vq_1+\vq_2|^{2n_3}  |\vk+\vq_1+\vq_2|^{2n_3'}},
\end{align}
which reduces to
\begin{align}
  \label{eq:I33WithXiN}
  I_{33}(k,\valpha,\vbeta) =& \int\d^3\vr\,e^{i\vk\cdot\vr}
\,\Xi^{(1)}_{n_3'n_3}(\vr,\vk)
\,\Xi^{(1)}_{n_1n_1'}(\vr+\valpha,\vk)\non\\
&\times \Xi^{(1)}_{n_2n_2'}(\vr+\vbeta,\vk).
\end{align}

\begin{figure*}[t!]
\includegraphics[scale=0.75]{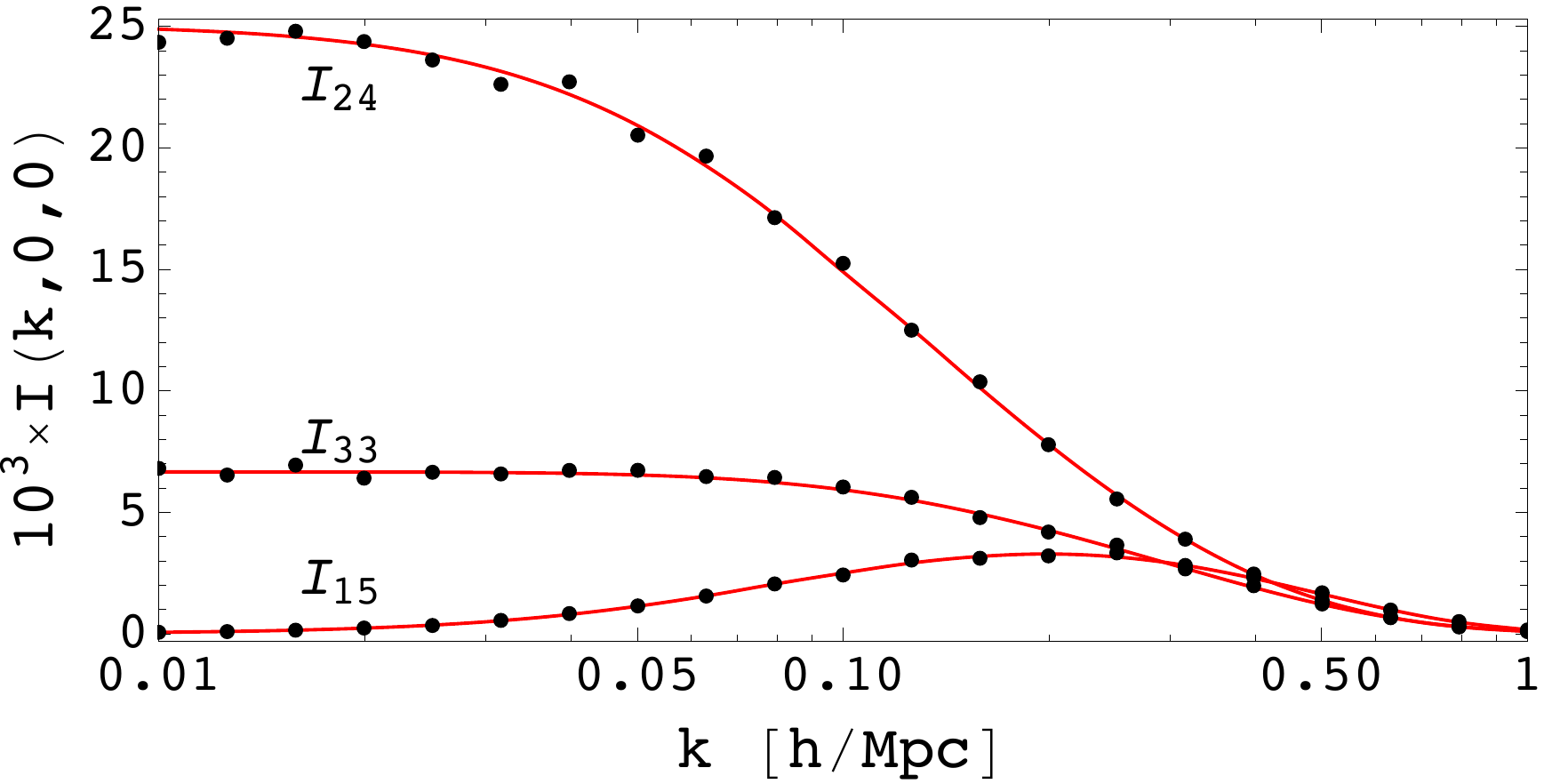}
\caption{Numerical results for two-loop integrals $I_{15}$, $I_{24}$ and $I_{33}$ defined in 
Eqs.~\eq{I15def}, \eq{I24def} and \eq{I33def}
for the special case where $\valpha=\vbeta =0$ and 
the linear power spectrum has the simple form $P_\lin=k^2 \exp (-k^2)$. 
Using the radial integrals in Eqs.~\eq{I15WithXiN}, 
\eq{I24WithXiN} and \eq{I33WithXiN} together with
\eqq{QNIntegralExpanded} (red lines) is compared against the conventionally
used direct Monte Carlo evaluation
of two-loop integrals using 
the \textsf{Cuba} \cite{cuba} library  (black points). 
As mentioned in the text, denominators in the $I_{15}$, $I_{24}$ and $I_{33}$  terms have been 
extended, $q^2 \to q^2 +\epsilon$ with $\epsilon=0.001$, in order order to avoid singular points. 
} 
\label{fig:01}
\end{figure*}

\subsection{Evaluating $\Xi^{(n)}$ integrals}

It remains to compute the $\Xi^{(n)}$ integrals defined in Eqs.~\eq{Xi0Def}, 
\eq{Xi1Def} and \eq{Xi2Def}. 
Similarly to calculations in the rest of the paper (expanding all terms in spherical harmonics or using Eqs.~C4, C6 and C7 of \cite{MarcelZvonimirPat1603}) we find
\begin{align}
&\Xi^{(N)}_{nn'}(\vr,\vk)\non\\ 
&\quad=\, \sum_L (-i)^L (2L+1)
\mathsf{P}_L(\hat\vr\cdot\hat\vk)
Q^{(N)}_{nn',L}(r,k),
  \label{eq:XiNExpansion}
\end{align}
where $\mathsf{P}_L$ are Legendre polynomials.

The $Q^{(N)}$ coefficients in \eqq{XiNExpansion} can be computed
in two alternative ways. One way is
\begin{align}
  \label{eq:QNIntegral}
Q^{(N)}_{nn',L}(r,k)
= &\,
4\pi\int_0^\infty \d s\,s^2 j_L(ks)
R^{(N)}_{n'}(s)
\int_0^\infty\frac{\d q}{2\pi^2}\,
\non\\
&\times 
q^{2-2n}
j_L(qs)j_L(qr)
\tilde R^{(N)}(q),
\end{align}
where $\tilde R^{(0)}(q)=1$, $\tilde R^{(1,2)}(q)=P_\lin(q)$,
$R^{(0)}_{n'}(s) = R^{(1)}_{n'}(s) =  \Xi^0_{-2n'}(s)$ (see
\eqq{FTofInvLaplacian} and \cite{MarcelZvonimirPat1603}),
and $R^{(2)}_{n'}(s) \equiv  \xi^0_{-2n'}(s)$.
To compute the two-dimensional radial integral in \eqq{QNIntegral}, 
we need to integrate over $q$ for 
every value of $s$ and $r$, and then perform 
a 1D Hankel transform for every value of $r$ and $k$ to evaluate
the integral over $s$.
This procedure is computationally much
more expensive than the one-dimensional Hankel transforms in the 
rest of the paper, but should still be relatively fast compared
to the commonly used five-dimensional integrations, noting
also that formally related integrals have been successfully computed 
in another context in \cite{Slepian:2015qza,Slepian1607}.

An alternative, potentially faster way of computing the $Q^{(N)}$ coefficients in \eqq{XiNExpansion} follows by first expanding
$1/|\vk + \vq|^{2n'}$ in case of $\Xi^{(0)}$
and $\Xi^{(1)}$ or $P_\lin(|\vk + \vq|)/|\vk + \vq|^{2n'}$ 
in case of $\Xi^{(2)}$ in Legendre polynomials $\mathsf{P}_L(\hat\vk\cdot\hat\vq)$.
This gives for example
\begin{align}
&Q^{(2)}_{nn',L}(r,k) \non\\
&\;= \frac{(-1)^L}{(2L+1)}
\int\frac{\d q}{2\pi^2}\, q^{2-2n}\, j_L(qr)
 a^{(2)}_{n',L}(k,q) P_\lin(q),
\label{eq:QNIntegralExpanded}
\end{align}
where $a^{(2)}_{n',L}(k,q)$ are coefficients in the Legendre expansion above.
For $Q^{(1)}$, we obtain the same expression but involving coefficients $a^{(1)}$ that follow from expanding $1/|\vk + \vq|^{2n'}$.
$Q^{(0)}$ can be obtained in the same way by omitting $P_\lin(q)$ in \eqq{QNIntegralExpanded}.

Finally, the expressions for $\Xi^{(N)}$ 
from \eqq{XiNExpansion}
are collected in the $I_n$ integrals in Eqs.~\eq{I15WithXiN}, 
\eq{I24WithXiN} and \eq{I33WithXiN}. 
The result then simplifies by expanding $\exp(i\vk\cdot\vr)$ in Legendre polynomials and using
\eqq{IntegrateFourLegendres} for the integral over four Legendre polynomials
(assuming the special case $\valpha=\vbeta=0$ with trivial numerator).

In Fig.~\ref{fig:01} we present the test results for the 
integrals $I_{15}(k,0,0)$, $I_{33}(k,0,0)$ and $I_{24}(k,0,0)$,
with $n_i=n_i'=1$ and each term in the denominators extended 
by an infinitesimally small
$\epsilon$ contribution in order to remove potential singular 
points, e.g.~we have $q^2 \to q^2 +\epsilon$ and similar for the 
rest of the terms. For realistic $P_{15}$, $P_{33}$ and $P_{24}$
terms, $\epsilon$ can be taken to zero.
We compare numerically computed results using Monte Carlo
\cite{cuba}
integration (points in Fig.~\ref{fig:01}) with the results computed with methods presented above 
using the $\Xi^{(N)}_{nn'}$ functions and \eqq{QNIntegralExpanded} (lines in Fig.~\ref{fig:01}), finding good overall agreement. 

The latter method is computationally much less expensive since 
we are reducing the five-dimensional integration to
1D integrals (computation of Legendre coefficients $a^{(N)}$)
and two sequences of consecutive Hankel transforms (one to obtain $Q^{(N)}$ using e.g.~\eqq{QNIntegralExpanded} and another one to evaluate e.g.~\eqq{I15WithXiN}).
Alternatively, $Q^{(N)}$
can be computed using the 2D integration in Eq.~\eqref{eq:QNIntegral}.

Both methods rely on the (infinite) summation over the multipole number 
$L$, but in practice this converges rapidly for $k<1\,h\mathrm{Mpc}^{-1}$,
so that the summation can be truncated, at least for the simple
test case with trivial numerator
and exponentially decaying linear power spectrum considered in Fig.~\ref{fig:01}.
Future numerical work is required to check how well this approach works
with nontrivial numerators and realistic linear power spectrum.

\section{Applicability, extensions and discussion}
\label{se:Comments}

\subsection{Functional form of the linear input power spectrum} 
\label{se:Pkshape}
Our equations are formally correct for an arbitrary linear input power
spectrum because no step of the derivations makes any assumption about
the shape of the input power spectrum.  At a practical level, the 1D
integrals are 1D Hankel transforms that are nontrivial to evaluate
numerically because of highly oscillatory spherical Bessel functions
in integrands.  Fortunately, these integrals can be evaluated robustly
and efficiently as 1D FFTs using the \textsf{FFTLog} library
\cite{hamiltonfftlog}.  This does impose a weak restriction on the
shape of the input power spectrum in the sense that it needs to be
 stored on a
discrete 1D grid so that no features finer than the grid resolution
can be represented.
However, we can use an extremely high resolution for this grid, because we
only need to perform one-dimensional FFTs on it, which are extremely
fast.  This resolution is more than sufficient to resolve features in
the power spectrum such as BAO wiggles. 

To see this more explicitly, note that the peaks and troughs of the
BAO wiggles in the power spectrum have a typical width of $\Delta
k\sim 0.02\,h\mathrm{Mpc}^{-1}$. 
For our 1D FFTs, we can easily use 10,000 grid points that are 
logarithmically
spaced in $10^{-5}\,h\mathrm{Mpc}^{-1}<k<100\,h\mathrm{Mpc}^{-1}$. 
This then gives more than 100 grid points
between $k=0.09\,h\mathrm{Mpc}^{-1}$ and
$k=0.11\,h\mathrm{Mpc}^{-1}$, and 40 grid points between 
$k=0.29\,h\mathrm{Mpc}^{-1}$ and
$k=0.31\,h\mathrm{Mpc}^{-1}$.  Every peak and trough of the BAO wiggles
can therefore easily be represented with dozens of grid points each,
which should indeed be sufficient to accurately model these BAO features.

Another potential restriction is that the FFTs used by
\textsf{FFTLog} may introduce ringing in the Hankel transforms.  In
\cite{MarcelZvonimirPat1603} we suppressed this by extrapolating the
linear input power spectrum with power laws at extremely large scales
$k\lesssim 10^{-5}\,h\mathrm{Mpc}^{-1}$ and at extremely small scales
$k\gtrsim 100\,h\mathrm{Mpc}^{-1}$, which do ultimately not contribute
significantly to the power spectrum on scales of practical interest
for cosmology.  This is therefore just a numerical trick to avoid
ringing and should not restrict the applicability of our method in
practice (also noting that numerical \textsf{FFT-PT} results for the 
1-loop power spectrum
were shown to agree with Monte-Carlo integrals at the 
$10^{-5}$ level \cite{MarcelZvonimirPat1603}).

\subsection{Number of terms}
For 2-loop power spectrum contributions with at most one inverse Laplacian,
our final expressions
involve only a finite number of terms that need to be summed up.  This
follows from the fact that the angular structure of the perturbative
$F_n$ and $G_n$ kernels does not go beyond a maximum multipole, which in turn
follows from the structure of the equations of motion for the DM
fluid.  Concretly, the $F_2$ and $G_2$ kernels involve at most
quadrupole terms like $\mathsf{P}_2(\hat\vq_i\cdot\hat\vq_j)$, and the
recursion relations \eq{RecursionRelns} imply that in general $F_n$
and $G_n$ involve at most $\mathsf{P}_n(\hat\vq_i\cdot\hat\vq_j)$.
Therefore, 1-5 terms involve only $\ell\le 5$, while 2-4 and 3-3 terms
involve only $\ell\le 6$.  
The total number of terms may still be significant.  While we have not
checked if this would be an issue in practice, we expect that even a
potentially large number of 1D FFTs should be faster than performing
five-dimensional Monte-Carlo integrals for every $k$ of interest.  The
number of FFTs can be reduced by exploiting symmetries to avoid
computing the same terms multiple times.  Since only a limited number
of $\xi^\ell_n$ are needed for all 2-loop integrals, some speedup
should also follow by computing all of them with 1D Hankel transforms
from a given linear power spectrum and storing them in memory, which
is trivial because they are defined on a 1D grid.

For contributions with two or more multiple inverse Laplacians, we followed
another approach that involves series of infinitely many terms. 
For the special cases considered in Section~\ref{se:P2LoopManyInvLaplacians},
we found that they can be truncated after a few terms. While we expect
this to be also the case in full generality, further work is required to check this.

\subsection{Potential infrared divergences}
While we showed how to evaluate 
2-loop power spectrum contributions using low-dimensional radial integrals, 
an important future step is to actually implement
this and test numerical performance in practice.  One potential issue
might be that we do not explicitly cancel the sensitivity of
individual contributions to very large-scale infrared (IR) modes within
integrands \cite{2014JCAP...07..056C}, but instead we currently rely
on accurate cancellations between fully integrated contributions.
While this seems problematic for certain power law initial power
spectra (scaling universes), it is less problematic for $\Lambda$CDM
initial power spectra that scale as $k^1$ on large scales $k\lesssim
k_\mathrm{eq}$.  Since the individual contributions should be accurate
to machine level precision if evaluated with FFTs, it should be
possible to control cancellations of large terms, but this needs to be
checked numerically.  If this poses problems in practice, an
alternative would be to modify the scheme so that IR sensitivity is
cancelled at an earlier stage of the algorithm.  
In this context it is also worth noting that our reformulation of 2-loop
integrals is by no means unique, and other reformulations may be more suitable
for numerical evaluations
(also see \cite{MarcelZvonimirPat1603}, where vector identities were used to reformulate some 1-loop results).

\subsection{\texorpdfstring{LPT and beyond $\Lambda$CDM}{LPT and beyond LCDM}}

Throughout our paper we have worked with the standard time-independent
perturbative $F_n$ kernels in SPT.  In LPT, the corresponding kernels
have slightly different coefficients but involve the same types of
terms when computing cumulants of the displacement field perturbatively
(e.g.~\cite{bernardeauReview,Matsubara:2007wj,1995MNRAS.276..115C,1996MNRAS.282..455C,rampf1203SPTvsLPTkernels, 2014PhRvD..90d3537M, Marcel1508}).
Our results can therefore straightforwardly be applied to 2-loop
integrals if LPT is evaluated order by order, simply by changing
coefficients (see \cite{MarcelZvonimirPat1603} for examples of this
for 1-loop integrals in LPT). Mapping from the displacement
cumulants to the density power spectrum in LPT can involve a second layer of 
computational complexity, but this can again be reduced to 
spherical Hankel transforms \cite{Vlah:2015sea, 2015JCAP...09..014V}.

The form of the perturbative $F_n$ kernels is strictly speaking only
valid in an Einstein-de Sitter (EdS) universe.  In other cosmologies
the kernels can be time-dependent. The effect of this on the 1-loop
matter power spectrum is typically at a sub-percent level
\cite{Takahashi0806,bernardeauReview}, but can reach $1\%$ or more
when also considering momentum statistics that
are relevant for redshift space distortions \cite{FasielloVlah1604}.
It would be interesting to test this approximation at the 2-loop
level. While this goes beyond the scope of this paper, our formalism
should still apply to the general cosmologies for which
Ref.~\cite{FasielloVlah1604} derived separable perturbative kernels.

\subsection{Halo bias}
\label{se:bias}
Tracers of the large-scale DM distribution 
such as halos or galaxies are typically biased
with respect to the DM.  The relation between halos and DM is often
modeled with a bias relation of the form \cite{McDonaldRoy2009,BiasReview2016}
\begin{align}
  \label{eq:25}
\delta_h(\vx)=b_1\delta_m(\vx)+b_2\delta_m^2(\vx)+b_{s2}s^2_m(\vx)+
b_{3}\delta^3_m(\vx)+\cdots,
\end{align}
where $s^2$ is the square of the DM tidal tensor, and we did not write down
velocity bias and potential other biases.  One way to include
this in perturbative models is to modify the perturbative $F_n$
kernels such that they relate the nonlinear halo density $\delta_h$ to the
linear DM density $\delta_1$, i.e.
\begin{align}
  \label{eq:27}
  \delta_h(\vk) = \;& \tilde F_1(\vk)\delta_1(\vk)\non\\
& + \int_\vq \tilde
  F_2(\vq,\vk-\vq)\delta_1(\vq)\delta_1(\vk-\vq)
+\cdots.
\end{align}
For example for the above simple bias relation the modified kernels
would be $\tilde F_1=b_1$ and 
\begin{align}
  \label{eq:26}
  \tilde F_2(\vq,\vp)  \;=\; &\left(\frac{17}{21}b_1+b_2\right)
\,+\, \frac{b_1}{2}\left(\frac{q}{p}+\frac{p}{q}\right)\hat\vq\cdot\hat\vp\non\\
& \,+\, \left(\frac{4}{21}b_1+b_{s^2}\right)
\frac{3}{2}\left((\hat\vq\cdot\hat\vp)^2-\frac{1}{3}\right).
\end{align}
This only changes coefficients, e.g.~from $17/21$ to $17/21\,b_1+b_2$,
without changing the structure of the terms contributing to the
kernels.  2-loop corrections to the halo power spectrum can
therefore be evaluated in the same way as for the DM power spectrum if
the modified coefficients of the halo $\tilde F_n$ kernels are used.
The $G_n$ velocity kernels should be modified in a similar way.

\subsection{Redshift space distortions}
\label{se:rsd}

Redshift space distortions (RSD) \cite{Jackson1972, Kaiser1987, Hamilton1998} emerge due to the fact that 
we observe redshifts of galaxies and not directly their positions. 
The position inferred from the observed redshift is 
distorted by the peculiar velocity and the comoving
redshift-space coordinate for a galaxy is given by
\begin{align}
  \vec s = \vec x +\hat \vz \frac{u_\parallel}{\mathcal H}
\end{align}
where $\hat\vz $ is the unit vector along the line of sight, 
and $u_\parallel$ is the comoving velocity parallel to
the line of sight.

There have been several approaches computing the  
RSD effects within PT \cite{Scoccimarro2004, Jeong2006, Taruya2010, Challinor2011, 
Reid2011, Seljak2011, Vlah2012, Vlah2013}. Even though initial computational routes of these 
approaches might seem rather different, the results are 
equivalent, as expected (assuming the same perturbative order, 
approximations and 
resummations in each of the approaches).

In the distribution function (DF) approach \cite{Seljak2011, Okumura2012, Vlah2012, Okumura2012v2, Vlah2013}
the overdensity in redshift space is given as the decomposition 
\begin{align}
    \delta_s(\vec{k})=\sum_{L=0}\frac{1}{L!}\left(\frac{ik_\parallel}{\mathcal{H}}\right)^LT^L_\parallel(\vec{k}),
\end{align}
were $T^L_\parallel(\vec{k})$ is the Fourier transform of velocity moments 
$T^L_\parallel(\vec{x})=(1+\delta(\vec{x}))v^L_\parallel(\vec{x})$.
It follows that the redshift space power spectrum in the plane-parallel approximation 
can be written as 
\begin{align}
    P^{ss}(\vec {k})=\sum_{\substack{L=0\\L'=0}}\frac{(-1)^{L'}}{L!\,L'!}\left(\frac{ik_\parallel}{\mathcal{H}}\right)^{L+L'}P_{LL'}(\vec{k}),
\end{align}
where $P_{LL'}(\vec {k})=\left\langle T^L_\parallel(\vec {k})\right.\left|T^{*L'}_\parallel(\vec{k}')\right\rangle'$ 
are the correlations of the different velocity moments. Using rotational symmetry, as shown in \cite{Seljak2011, Vlah2012},
each of the $P_{LL'}$ spectra can be further decomposed in the form
\begin{align}
 P_{LL'}(\vec {k})=\sum_{\substack{l=L,L-2,\ldots \\ l'=L',L'-2,\ldots\\ m=0\ldots l}}P^{L,L',m}_{l,l'}(k)\,\mathsf{P}^m_l(\mu)\,\mathsf{P}^m_{l'}(\mu),
\end{align}
where $\mathsf{P}^m_l(\mu)$ are the associated Legendre polynomials, and $\mu = \hat \vz \cdot \hat \vk$.
It it important to note that the decomposed spectra $P^{L,L',m}_{l,l'}$ depend 
only on the magnitude $k$ of the wavevector $\vec k$. 
Also note that the decomposition above gives
a finite number of terms for each $L$ and $L'$.
Explicit PT expression for all the 1-loop $P^{L,L',m}_{l,l'}$ 
contributions are given
in Ref.~\cite{Vlah2012}.
They can be constructed from $I_{nm}(k)$ and $J_{nm}(k)$ 
expressions given in Appendix D of Ref.~\cite{Vlah2012}. It is clear that the
\textsf{FFT-PT}
method used for the fast computation of the $P_{22}$ and $P_{13}$ 
1-loop contributions from Ref.~\cite{MarcelZvonimirPat1603} is straightforwardly applicable 
to the integrals $I_{nm}(k)$ (convolution type integrals similar to $P_{22}$) 
and $J_{nm}(k)$ (propagator type integrals similar to $P_{13}$).

It is important to note that the decomposition of the RSD effect into the $P^{L,L',m}_{l,l'}(k)$ 
spectra does not rely on PT and is valid up to all orders. So analogous
expressions as presented up to one loop in~\cite{Vlah2012} can be computed
up to two loop.  For these correlations the methods presented in this paper would be fully 
applicable.

As mentioned, an advantage of the DF approach lies in the use of 
rotational symmetries to determine the angular structure of RSD 
correlators valid regardless of the PT order. 
One-loop RSD power spectrum results obtained in some of the other 
references~\cite{Scoccimarro2004, Jeong2006, Taruya2010, Challinor2011}~\footnote{Note that some of the references use stronger approximations, dropping 
some terms compared to the DF approach. 
We are interested here in comparing 
the PT structure (especially angular) of these results where they overlap,
disregarding the differences.}
reduce after explicit calculation to the same angular structure, as expected, 
finally reaching the same conclusion, albeit, in a less transparent way. 
Our treatment of RSD corrections to the 2-loop power spectrum is therefore not restricted to the DF approach but applies to all the other RSD modeling approaches mentioned above.
This discussion should also be valid at higher orders.
  
Similar conclusions (keeping in mind Section~\ref{se:bias}) hold for  biased tracers in redshift space where the explicit 
decompositions using the DF approach can be found in \cite{Vlah2013}.
From the above it follows that equivalent conclusions hold also for velocity 
statistics (pairwise velocity and pairwise dispersion) in real and redshift space 
and the explicit DF decomposition presented in \cite{Okumura2014}.

As mentioned above, our methods for rapid loop computations can
also be performed in LPT (for explicit 1-loop expressions see appendices in e.g.~\cite{Matsubara:2007wj}). 
For LPT models of RSD effects in addition to the biasing effects see e.g.~\cite{Carlson2013, Vlah:2016}.

\section{Conclusions}
\label{se:conclusions}

Pushing models of the large-scale structure of the universe to 
nonlinear
scales is a challenging problem in
cosmology.  An extensively studied approach to this 
is perturbation theory.
Unfortunately, perturbative corrections come in the form of high-dimensional loop
integrals that are cumbersome to evaluate. 
For example, the 2-loop power spectrum involves
five-dimensional integrals at every wavenumber of interest.

Generalizing previous work on the 1-loop matter power spectrum
\cite{MarcelZvonimirPat1603,OSUloops1603}, we show in this paper how
 2-loop corrections to the density
power spectrum in Eulerian standard perturbation theory can be
rewritten so that they involve only low-dimensional radial integrals.
In absence of multiple inverse Laplacians, these take the form of 
one-dimensional Hankel transforms that can be evaluated
very efficiently with one-dimensional
FFTs using \textsf{FFTLog} \cite{hamiltonfftlog}.  
Contributions arising from multiple inverse Laplacians seem to require a sequence of low-dimensional radial integrals, which are computationally more challenging but may still be faster than five-dimensional integrations (see Section~\ref{se:P2LoopManyInvLaplacians}).

One specific use case of this \FFPT~method is the possibility to speed up Monte-Carlo
chains when fitting cosmological parameters from LSS observations.
More generally, the fast expressions can be useful for anyone working
with the 2-loop power spectrum or higher-order loop integrals in general.

Our reformulation of 2-loop power spectrum integrals is based on avoiding convolution integrals by repeatedly changing between Fourier and position space, integrating over orientations, and performing the remaining radial integrals using one-dimensional FFTs.
This
is very general in
the sense that it does not assume a specific shape for the linear
input power spectrum.
This, in turn, is important to accurately model the imprint of baryonic acoustic
oscillations on LSS 2-point statistics, which is arguably the most
pristine cosmological signal measured with high precision from
modern surveys.  

The result that three-dimensional loop integrals can be reduced to 
one-dimensional 
radial integrals is not a coincidence,
but can be understood from the fact
that structure formation only depends on distances 
between objects if we assume statistical isotropy and homogeneity
and the standard fluid equations of motion with their 
standard perturbative solution (also see 
\cite{MarcelZvonimirPat1603}).

We show how the same method can be applied to the 2-loop power spectrum
of halos or any other biased tracer of the dark matter with known
bias relation.
Redshift space distortions can also be handled with this method.
This is straightforward to see for the distribution
function approach to model redshift space distortions but also applies to many other RSD modeling approaches (see Section~\ref{se:rsd}).
Our method should also apply to Lagrangian space models
as shown for the 1-loop case in \cite{MarcelZvonimirPat1603}.
For the special, presumably only academically interesting case of scaling universes with perfect power law
initial power spectrum, the one-dimensional FFTs 
can be evaluated analytically so
that all 2-loop power spectrum contributions reduce to simple power laws
(see Appendix~\ref{se:ScalingUni}).

In the future, it would be interesting to 
numerically implement the fast 2-loop expressions presented in our
paper, extending the 1-loop 
implementations of \cite{MarcelZvonimirPat1603,OSUloops1603}.
It would also be useful to
include
effective field theory corrections and generalize 
the method to higher-order statistics like
the bispectrum or trispectrum.  These possible directions of future investigation
seem worthwhile pursuing
given the impressive amount of upcoming data from a number of planned
LSS surveys in the near future and the need to analyze and model these
observations beyond the linear regime to maximize their science returns.

\section*{Acknowledgments}
We thank Pat McDonald, Tobias Baldauf, Simon Foreman, Marko Simonovich
and Matias Zaldarriaga
for very useful discussions related
to this work.
We also thank Simon Foreman for comments on the manuscript.
Z.V.~is supported in part by the U.S.~Department of Energy contract to SLAC no.~DE-AC02-76SF00515.

\appendix

\section{Perturbative expansion and gravity kernels}
\label{se:SPTreview}

This section provides a brief overview of the perturbative approach to solve the equations of motion in Eulerian standard perturbation theory (see \cite{bernardeauReview} for a review).

We start form the standard ansatz for the expansion of density and velocity divergence field
\begin{align}
\df(\vec k, \tau) & = \sum_{n=0}^{\infty} \df^{(n)}(\vec k, \tau),\\
\tf(\vec k, \tau) & = -f(\tau)\mathcal H (\tau) \sum_{n=0}^{\infty} \tf^{(n)}(\vec k, \tau),
\end{align}
where we have for a given order
\begin{align}
 \df^{(n)}(\vec k, \tau) &= 
 \int_{\vec q_1 \ldots \vec q_n } 
 (2\pi)^3 \df^D (\vec k - \vec q_1 \ldots -\vec q_n) \non\\
&\qquad\times
 F^{(s)}_n(\vec q_1,\ldots, \vec q_n) \df_1 (\vec q_1, \tau)\ldots \df_1 (\vec q_n, \tau), \non\\
 \tf^{(n)}(\vec k, \tau) &= 
 \int_{\vec q_1 \ldots \vec q_n } 
 (2\pi)^3 \df^D (\vec k - \vec q_1 \ldots -\vec q_n) \non\\
&\qquad \times G^{(s)}_n(\vec q_1,\ldots, \vec q_n) \df_1 (\vec q_1, \tau)\ldots \df_1(\vec q_n, \tau).
\label{eq:PTfileds}
\end{align}
By definition the first order kernels are unity, i.e. $F_1^{(s)}=G_1^{(s)} =1$. Since the linear solution $\df_1$ 
is known, all higher order nonlinearities are incorporated in the kernels $F^{(s)}_n$ and $G^{(s)}_n$. 
The upper index $(s)$ denotes symmetrized kernels,
\begin{align}
F^{(s)}_n \lb \vec q_1 , \ldots , \vec q_n \rb & = \frac{1}{n!} \sum_\pi F_n \lb \pi \{ \vec q_1 , \ldots , \vec q_n \} \rb, \non\\
G^{(s)}_n \lb \vec q_1 , \ldots , \vec q_n \rb & = \frac{1}{n!} \sum_\pi G_n \lb \pi \{ \vec q_1 , \ldots , \vec q_n \} \rb .
\end{align}
Un-symmetrized kernels satisfy recursion relations that can be derived by substituting Eq.~\eqref{eq:PTfileds}
into the equations of motion Eq.~\eqref{eq:EulerEoM}.
These recursion relations are
\begin{align}
&F_n(\vec{q}_1,\dots,\vec{q}_n)
= 
\sum_{m=1}^{n-1}\frac{G_m(\vec{q}_1,\dots,\vec{q}_m)}{(2n+3) (n-1)}\non\\
&\;\times\bigg\{(2n+1)\frac{\vec{k}\cdot \vec{q}_{1\cdots m}}{q_{1\cdots m}^2}F_{n-m}(\vec{q}_{m+1},\dots,\vec{q}_n) \non\\
&\;\quad + \frac{k^2 \vec{q}_{1\cdots m}\cdot\vec{q}_{m+1\cdots n}}{q^2_{1\cdots m} 
q^2_{m+1\cdots n}} G_{n-m}(\vec{q}_{m+1},\dots,\vec{q}_n)\bigg\}
\label{eq:RecursionRelns}
\end{align}
and
\begin{align}
&G_n(\vec{q}_1,\dots,\vec{q}_n) = 
\sum_{m=1}^{n-1}\frac{G_m(\vec{q}_1,\dots,\vec{q}_m)}{(2n+3) (n-1)}\non\\
&\;
\times\bigg\{3 \frac{\vec{k}\cdot \vec{q}_{1\cdots m}}{q_{1\cdots m}^2}F_{n-m}(\vec{q}_{m+1},\dots,\vec{q}_n) \non\\
& \;\quad+ n \frac{k^2 \vec{q}_{1\cdots m}\cdot\vec{q}_{m+1\cdots n}}{q^2_{1\cdots m} 
q^2_{m+1\cdots n}} G_{n-m}(\vec{q}_{m+1},\dots,\vec{q}_n)\bigg\}
\end{align}
where we have introduced the notation $\vec{q}_{1\cdots m}=\vec{q}_{1}+\cdots+\vec{q}_m$ and 
$\vec{q}_{m+1\cdots n}=\vec{q}_{m+1}+\cdots+\vec{q}_n$.
Also $\vk=\vq_1+\cdots+\vq_n$ in the last two equations.

\section{Coupling factors}
\label{se:CouplingFactors}

This Appendix provides analytical expressions for the coupling factors $\mathcal{M}_n$ that arise from the angular structure of 2-loop integrands and are used throughout the paper.  
The first few examples are evaluated in Table~\ref{tab:M33}.

1-5 correlations without inverse Laplacians in \eqq{P15GeneralIntegralNoDenomi} are proportional to the coupling factor $\mathcal{M}_{0}$ defined by
\begin{align}
  \label{eq:M15}
\mathcal{M}_{0}(\ell_0,\ell_1,\ell_2)\equiv\sum_{\ell'=0}^{\mathrm{min}(\ell_0,\ell_1,\ell_2)} 
\alpha_{\ell_0\ell'}\alpha_{\ell_1\ell'}\alpha_{\ell_2\ell'}
\,(2\ell'+1).
\end{align}
The $\alpha_{\ell\ell'}$ coefficients follow from decomposing products between wavevectors in spherical harmonics using \eqq{ScalProdInYlms}, and are given by \cite{MarcelZvonimirPat1603,zvonimir1410}
\begin{align}
&\alpha_{\ell\ell'}  \;=\; \frac{1}{2} \int_{-1}^1\d\mu \,\mu^\ell\, \mathsf{P}_{\ell'}(\mu)\non\\
&=
  \begin{cases}
\frac{\ell!}{2^{(\ell-\ell')/2} \left[(\ell-\ell')/2\right]!\,
(\ell+\ell'+1)!!}, & \text{if 
$\;\ell\!\ge\! \ell'$ \& 
$\ell\!+\!\ell'$ even
}  ,\\
0, & \text{otherwise}.
  \end{cases}
  \label{eq:alphas}
\end{align}
These coefficients vanish if the second index $\ell'$ is greater than the first index $\ell$, which helps to render sums in the paper finite. 
They are normalized so that $\alpha_{00}=1$. 
The $\mathcal{M}_0$ coupling factor in \eqq{M15} is symmetric in its arguments, and all nonzero factors for $l_i\le 2$ are listed in Table~\ref{tab:M33}.

The coupling factor $\mathcal{M}_1$ between $(\ell_1,\ell_2,\ell_3)$ 
and $L$ that enters in \eqq{P24generalNonDenomi} is defined as
\begin{align}
  \label{eq:M24}
&  \mathcal{M}_{1}(\ell_1,\ell_2,\ell_3;L)\non\\
&\quad \equiv\,
(-1)^L (2L+1)
\sum\limits_{\ell_1'=0}^{\ell_1}
\alpha_{\ell_1\ell_1'}\,(2\ell_1'+1)
\non\\
&\qquad\quad\times\sum\limits_{\ell'=0}^{\mathrm{min}(\ell_2,\ell_3)}
\alpha_{\ell_2\ell'}\alpha_{\ell_3\ell'}
(2\ell'+1)
\left(\begin{matrix}
  L  & \ell_1' & \ell'\\
  0 & 0 & 0
\end{matrix}\right)^2
.
\end{align}
This involves a Wigner 3-j symbol, which imposes a triangle condition that implies $L\le \ell_1'+\ell'\le \ell_1+\mathrm{min}(\ell_2,\ell_3)$.
The range of allowed $L$ in \eqq{P24generalNonDenomi} is therefore finite. 
The coupling factor is symmetric under $\ell_2\leftrightarrow \ell_3$.
For $\ell_i\le 1$ the only nonzero couplings are shown in Table~\ref{tab:M33}.

In \eqq{P24GeneralFastCase1} we used the coupling factor $\mathcal{M}_{2}$ between $(\ell_1,\ell_2,\ell_3)$ and $(L,L')$, which is 
defined as
\begin{align}
  \label{eq:M24tildedef}
&\mathcal{M}_{2}(\ell_1,\ell_2,\ell_3; L,L')
\non\\
& \equiv
(-1)^{L+L'}
(2L\!+\!1)(2L'\!+\!1)
\sum_{\ell_1'=0}^{\ell_1}\sum_{\ell_2'=0}^{\ell_2}\sum_{\ell_3'=0}^{\ell_3} \alpha_{\ell_1\ell_1'}\alpha_{\ell_2\ell_2'}\alpha_{\ell_3\ell_3'}
\non\\
&\;\;\times
(2\ell_1'\!+\!1)(2\ell_2'\!+\!1)(2\ell_3'\!+\!1)
\bigg(\begin{matrix}
  \ell_1' & L' & \ell_2'\\
   0 & 0 & 0
\end{matrix}\bigg)^2
\bigg(\begin{matrix}
  \ell_2' & L & \ell_3'\\
   0 & 0 & 0
\end{matrix}\bigg)^2.
\end{align}
The 3-j symbols impose triangle conditions on $(L',\ell_1',\ell_2')$ and $(L,\ell_2',\ell_3')$, which make the sums over $L$ and $L'$ in \eqq{P24GeneralFastCase1} finite.
The coupling factor is symmetric under simultaneously changing $\ell_1\leftrightarrow \ell_3$ and $L\leftrightarrow L'$.
All nonzero couplings for $\ell_1\le\ell_2\le\ell_3\le 1$ are listed in Table~\ref{tab:M33}.

Finally, several expressions involve the coupling $\mathcal{M}_3$ between $(\ell_1,\ell_2,\ell_3)$ and $(L_1,L_2,L_3)$ defined by 
\begin{align}
&  \mathcal{M}_{3}(\ell_1,\ell_2,\ell_3;L_1,L_2,L_3) 
\non\\
& \quad\equiv\; 
\sum_{\ell_1'=0}^{\ell_1}\sum_{\ell_2'=0}^{\ell_2}\sum_{\ell_3'=0}^{\ell_3}
\alpha_{\ell_1\ell_1'}\alpha_{\ell_2\ell_2'}\alpha_{\ell_3\ell_3'}
\left\{\begin{matrix}
L_1 &  L_2 & L_3  \\
\ell_1' & \ell_2' & \ell_3' 
\end{matrix}\right\}'.
  \label{eq:M33def}
\end{align}
This involves a rescaled 6-j symbol defined in \eqq{modified6j}.
It
severely restricts the allowed values for $L_i$ so that e.g.~the sum on the
right hand side of \eqq{P33simple1} is finite.\footnote{In general, $\alpha$ coefficients
  enforce $\ell_i'\le \ell_i$, where $\ell_i$ denote exponents of
  scalar products in the integrand of \eqq{P33simple1}, and the 6-j
  symbol enforces four triangle conditions, restricting
  $|\ell_2'-\ell_3'|\le L_1\le \ell_2'+\ell_3'$, and similarly for
  $L_2$ and $L_3$, as well as $|L_2-L_3|\le L_1\le L_2+L_3$.
  Additionally, the modified 6-j symbol \eq{modified6j} enforces
  $L_1+L_2+L_3$, $L_1+\ell_2'+\ell_3'$, $L_2+\ell_1'+\ell_3'$ and
  $L_3+\ell_1'+\ell_2'$ to be even.  The coupling factor has the same
  symmetry properties as the 6-j symbol.  }

\section{2-4 correlations with inverse Laplacians and nontrivial angular dependence}
\label{app:P24WithInvLaplAndAnglesAppdx}

This Appendix shows how to evaluate nontrivial 2-4 correlations with inverse Laplacians and nontrivial angular dependence, providing details of the results summarized in Section~\ref{se:P24WithInvLaplAndAngles}.
Additional details on the derivation of these results will be provided in Appendix~\ref{se:P24details}.

\subsection{Most general form of 2-4 correlations}

As mentioned before, the loop correction to the SPT power spectrum generated by 2-4 correlations is
\begin{align}
  P_{24}(k) & = 24\int_{\vq_1\vq_2}  F^{(s)}_4(\vq_1,-\vq_1,\vq_2,\vk\!-\!\vq_2)\,F^{(s)}_2(\vq_2,\vk\!-\!\vq_2) \non\\
&\qquad \times P_\lin(q_1)\,P_\lin(q_2)\,P_\lin(|\vk\!-\!\vq_2|).
  \label{eq:P24}
\end{align}
The most general form of terms contributing to this is
\begin{align}
& \int_{\vq_1\vq_2}
\big[\hat\vq_2\cdot\widehat{(\vk\!-\!\vq_2)}\big]^{\ell_1} 
\,\big[\hat\vq_1\cdot\widehat{(\vk\!-\!\vq_2)}\big]^{\ell_2}
\,\left(\hat{\vq}_1\cdot\hat{\vq}_2\right)^{\ell_3} \non\\
& \quad\times\frac{q_1^{n_1}P_\lin(q_1)\,q_2^{n_2}P_\lin(q_2)\,
|\vk\!-\!\vq_2|^{n_3}
P_\lin(|\vk\!-\!\vq_2|)}{|s_1\vq_1 + s_2\vq_2 + s_3(\vk\!-\!\vq_2)|^{2}},
  \label{eq:P24general}
\end{align}
where the momenta entering the denominator or inverse Laplacian are parametrized by the parameters $s_1\in\{-1,0,1\}$ and $s_2,s_3\in\{0,1\}$.
The only cases not already covered by \eqq{P24generalNonDenomi} occur when
at least two of these $s_i$ parameters are nonzero.
The angular structure in \eqq{P24general} is sufficiently general because it accounts for all scalar products that can be formed between the arguments of $F_4$ and between the arguments of $F_2$ in \eqq{P24}. 

\subsection{Splitting in two nested 1-loop integrals}

We now show how to evaluate the most general form of 2-4 correlations given in \eqq{P24general} by splitting it into two nested 1-loop integrals as demonstrated for a simpler case in \eqq{P24FastNoAngleDep}.
We proceed in two steps:  First, we calculate the loop integral over the $\vq_1$ momentum that
connects the $F_4$ vertex to itself, corresponding to the red tadpole subdiagram in Fig.~\ref{fig:TwoLoopDiagrams}.
We then insert the result and compute the integral over the other
loop momentum $\vq_2$ that connects the $F_4$ and $F_2$ vertices, corresponding to the blue subdiagram in Fig.~\ref{fig:TwoLoopDiagrams}.

Explicitly, to evaluate the 2-loop integral of \eqq{P24general}, we write scalar products involving the tadpole momentum $\vq_1$ in terms of spherical harmonics, e.g.
\begin{align}
  \label{eq:ScalProdInYlms24}
  (\hat{\vq}_1\cdot\hat{\vq}_2)^{\ell_3} = 4\pi \sum_{\vl'_3}^{\ell_3} \alpha_{\ell_3\ell'_3}
Y_{\vl_3'}(\hat\vq_1)Y^*_{\vl_3'}(\hat\vq_2),
\end{align}
where we use the condensed notation $\vl=(\ell,m)$ and 
$\sum_{\vl'}^\ell=\sum_{\ell'=0}^\ell\sum_{m'=-\ell'}^{\ell'}$.
Generalizing \eqq{P24generalSimple1b}, the integral \eq{P24general} then splits into an outer $\vq_2$-integral
\begin{align}
&\int_{\vq_2\vq_1} 
\big[\hat\vq_2\cdot\widehat{(\vk\!-\!\vq_2)}\big]^{\ell_1} 
\big[\hat\vq_1\cdot\widehat{(\vk\!-\!\vq_2)}\big]^{\ell_2}
\left(\hat{\vq}_1\cdot\hat{\vq}_2\right)^{\ell_3}\non\\
&\times \frac{q_1^{n_1}P_\lin(q_1)\,q_2^{n_2}P_\lin(q_2)\,
|\vk\!-\!\vq_2|^{n_3}
P_\lin(|\vk\!-\!\vq_2|)}{|s_1\vq_1 + s_2\vq_2 + s_3(\vk\!-\!\vq_2)|^{2}}\non\\
&\;\, =
4\pi
\int_{\vq_2} 
\big[\hat\vq_2\!\cdot\!\widehat{(\vk\!-\!\vq_2)}\big]^{\ell_1} 
q_2^{n_2}P_\lin(q_2)\,
|\vk\!-\!\vq_2|^{n_3}
\non\\
&\quad\;\;\times 
P_\lin(|\vk\!-\!\vq_2|)
\sum_{\vl_2'}^{\ell_2}
\sum_{\vl_3'}^{\ell_3}
\alpha_{\ell_2\ell_2'}
Y^*_{\vl_2'}(\widehat{\vk\!-\!\vq_2})
\non\\
&\quad\;\;\times
P^{\vl_2'\vl_3's_1n_1}_{\mathrm{tadpole}}(s_2\vq_2+s_3(\vk\!-\!\vq_2))
\,\alpha_{\ell_3\ell_3'}
Y^*_{\vl_3'}(\hat\vq_2),
\label{eq:P24nestedStep1}
\end{align}
over an inner tadpole integral over $\vq_1$, 
\begin{align}
  \label{eq:P24tadpole}
  P^{\vl_2'\vl_3's_1n_1}_{\mathrm{tadpole}}(\vp)
\equiv 4\pi \int_{\vq_1}
Y_{\vl_2'}(\hat\vq_1)\,Y_{\vl_3'}(\hat\vq_1)
\, \frac{q_1^{n_1}P_\lin(q_1)}{|s_1\vq_1+\vp|^{2}},
\end{align}
which is evaluated at the momentum $\vp\equiv s_2\vq_2+s_3(\vk\!-\!\vq_2)$.
Splitting the 2-loop integral in these two nested 1-loop integrals is the main trick needed for evaluating $P_{24}$.  
The remaining procedure to evaluate the outer and inner tadpole 1-loop integrals in Eqs.~\eq{P24nestedStep1} and \eq{P24tadpole} is similar to Section~\ref{se:P24TrivialAngles} and \cite{MarcelZvonimirPat1603}, as we show next.

\subsection{Evaluating the tadpole 1-loop integral over $\vq_1$}
The only cases for the denominator in \eqq{P24general} that are not already covered by the separable case in \eqq{P24generalNonDenomi} are $(s_1,s_2,s_3)\in\{(\pm 1,1,0), (\pm 1,0,1), (\pm 1,1,1)\}$, corresponding to $\vp\in\{\vq_2,\vk-\vq_2,\vk\}$ and $s_1=\pm 1$.
We therefore only consider these cases in the following.
The different cases correspond to different couplings and fields on which the inverse Laplacian acts, generalizing \eqq{P24generalSimple1a} from Section~\ref{se:P24TrivialAngles}.

In each case, the inner tadpole integral \eq{P24tadpole} is a Fourier-space convolution that reduces to a position space product of $\Xi^0_{-2}(r)=(4\pi r)^{-1}$ defined in \eqq{FTofInvLaplacian} and the correlation $\xi^L_{n_1,s_1}(r)$ defined in \eqq{xidefWithParam}:
\begin{align}
  \label{eq:P24tadpoletmp}
 P^{\vl_2'\vl_3's_1n_1}_{\mathrm{tadpole}}(\vp)
= 4\pi\sum_{L=0}^{\infty}\sum_{M=-L}^L (-1)^{L} \,
\mathcal{G}_{\vl_2'\vl_3'\VL}\,
Y_{\VL}^*(\hat\vp)\,
\mathcal{P}^{L}_{n_1,s_1}(p)
\end{align}
with 
\begin{align}
  \label{eq:CalPforP24}
  \mathcal{P}^{L}_{n_1,s_1}(p) \equiv \int_0^\infty\d r\,r\,
 j_{L}(pr)\,\xi^{L}_{n_1,s_1}(r).
\end{align}
This follows by integrating over orientations of the tadpole momentum $\vq_1$ and expanding in multipoles of the momentum $\vp$.
For $\ell_2'=\ell_3'=n_1=0$, $s_1=1$ and $\vp=\vq_2$, we recover the simpler result of \eqq{PtadpoleNoAngle} because $\mathcal{G}_{\mathbf{0}\mathbf{0}\mathbf{0}}=Y^*_{\mathbf{0}}(\hat\vp)=(4\pi)^{-1/2}$.

To proceed with the evaluation of the outer integral over $\vq_2$ in \eqq{P24nestedStep1}, we consider the cases $\vp\in\{\vq_2,\vk\!-\!\vq_2,\vk\}$ separately.

\begin{widetext}
\subsection{Evaluating the outer 1-loop integral: Case 1}

We start with the case $\vp=\vq_2$ in \eqq{P24tadpole}, i.e.~$(s_1,s_2,s_3)=(\pm 1,1,0)$.
As shown in detail in Appendix~\ref{se:P24details}, using \eqq{P24tadpoletmp} in \eqq{P24nestedStep1}, performing angular integrations, and exploiting orthogonality relations of Wigner 3-j symbols leads to
\begin{empheq}[box=\fbox]{align}
&\int\frac{\d\Omega_{\hat\vk}}{4\pi}
\int_{\vq_1\vq_2} 
\big[\hat\vq_2\!\cdot\!\widehat{(\vk\!-\!\vq_2)}\big]^{\ell_1} 
\big[\hat\vq_1\!\cdot\!\widehat{(\vk\!-\!\vq_2)}\big]^{\ell_2}
\left(\hat{\vq}_1\!\cdot\!\hat{\vq}_2\right)^{\ell_3}
\frac{q_1^{n_1}P_\lin(q_1)\,q_2^{n_2}P_\lin(q_2)\,
|\vk\!-\!\vq_2|^{n_3}
P_\lin(|\vk\!-\!\vq_2|)}{|s_1\vq_1 + \vq_2|^{2}}\non\\
\label{eq:P24GeneralFastCase1}
& \quad = 
4\pi \int_0^\infty\d r\,r^2 j_0(kr)
\sum_{L=0}^{\ell_2+\ell_3}
\sum_{L'=0}^{\ell_1+\ell_2}
\mathcal{M}_{2}(\ell_1,\ell_2,\ell_3;L,L')
\,\mathcal{T}^{s_1n_1n_2}_{LL'}(r)\,
\xi^{L'}_{n_3}(r).
\end{empheq}
This is a 1D Hankel transform of a sum of position space products of 4-point like correlations $\mathcal{T}(r)$ and linear correlation functions $\xi(r)$. 
The former is a generalization of \eqq{TauSimple}, given by a 1D Hankel transform of the Fourier space product of the linear power spectrum and the transformed linear correlation function  $\mathcal{P}$ from \eqq{CalPforP24}:
\begin{align}
  \label{eq:Tau24General}
\mathcal{T}^{s_1n_1n_2}_{LL'}(r) =
  \int_0^{\infty}\frac{\d q_2}{2\pi^2}\, q_2^{n_2+2}j_{L'}(q_2r)P_\lin(q_2)
\mathcal{P}^{L}_{n_1,s_1}(q_2).
\end{align}

The coupling factor $\mathcal{M}_{2}$ is defined by \eqq{M24tildedef}.
In the simple special case $\ell_1=\ell_2=\ell_3=0$, the only non-vanishing coupling is $\mathcal{M}_{2}(0,0,0;0,0)=1$, so that the general result of \eqq{P24GeneralFastCase1} simplifies and we recover the simple result of \eqq{P24FastNoAngleDep} derived earlier (for $s_1=1$).

The case $\vp=\vk-\vq_2$ in \eqq{P24tadpole} with $(s_1,s_2,s_3)=(\pm 1,0,1)$ follows by symmetry, noting that the result above does not change if we change the integration variable from $\vq_2$ to $\vk-\vq_2$.
Indeed, if we change the denominator on the left-hand side of \eqq{P24GeneralFastCase1} to $|s_1\vq_1+\vk-\vq_2|^{-2}$, the right-hand side follows by 
relabelling $\ell_2\leftrightarrow \ell_3$ and $n_2\leftrightarrow n_3$.

\subsection{Evaluating the outer 1-loop integral: Case 2}
 The last case is $\vp=\vk$ in \eqq{P24tadpole}, i.e.~$(s_2,s_3)=(1,1)$.
This case is related to the following contraction of the inverse Laplacian acting on a cubic field:
\begin{align}
  \label{eq:6}
&\int\d^3\vr\,e^{i\vk\cdot\vr}\,
{
\contraction[2.25ex]{\big\la}{\delta_1}{(\vx)\delta_1(\vx)\,\delta_1(\vx')\nabla^{-2}\big[\delta_1(\vx')\delta_1(\vx')}{\delta_1}
\contraction[1.5ex]{\big\la\delta_1(\vx)}{\delta_1}{(\vx)\,\delta_1(\vx')\nabla^{-2}\big[\delta_1(\vx')}{\delta_1}
\contraction[0.75ex]{\big\la\delta_1(\vx)\delta_1(\vx)\,}{\delta_1}{(\vx')\nabla^{-2}\big[}{\delta_1}
\big\la\delta_1(\vx)\delta_1(\vx)\,\delta_1(\vx')\nabla^{-2}\big[\delta_1(\vx')\delta_1(\vx')\delta_1(\vx')\big]\big\ra
}
=
-\int_{\vq_1\vq_2}\frac{P_\lin(q_1)P_\lin(\vq_2)P_\lin(\vk-\vq_2)}{|\vq_1+\vk|^2}.
\end{align}
The fully general case additionally contains scalar products between wavevectors. It can be reduced to (see endnote \footnote{Similarly to the calculation in Appendix~\ref{se:P24details}, angular integrations lead to four Gaunt integrals. 
Due to slightly different couplings the sum over $m$'s yields a 6-j symbol, leading to the coupling factor $\mathcal{M}_{3}$.
})
\begin{empheq}[box=\fbox]{align}
&\int\frac{\d\Omega_{\hat\vk}}{4\pi}\int_{\vq_1\vq_2} 
\big[\hat\vq_2\cdot\widehat{(\vk\!-\!\vq_2)}\big]^{\ell_1} 
\,\big[\hat\vq_1\cdot\widehat{(\vk\!-\!\vq_2)}\big]^{\ell_2}
\,\left(\hat{\vq}_1\cdot\hat{\vq}_2\right)^{\ell_3} 
\frac{q_1^{n_1}P_\lin(q_1)\,
q_2^{n_2}P_\lin(q_2)\,
|\vk\!-\!\vq_2|^{n_3}P_\lin(|\vk\!-\!\vq_2|)
}
{|s_1\vq_1 + \vk|^{2}}
\non\\
&\;=\,
4\pi\sum_{L=0}^{\ell_2+\ell_3} \mathcal{P}^L_{n_1,s_1}(k)
 \int_0^\infty\d r\,r^2 j_L(kr) 
\sum_{L_2=0}^{\ell_1+\ell_3}
\sum_{L_3=0}^{\ell_1+\ell_2}
\mathcal{M}_{3}(\ell_1,\ell_2,\ell_3;L,L_2,L_3)\;
\xi^{L_2}_{n_2}(r)\,\xi^{L_3}_{n_3}(r)
.
\label{eq:P24GeneralFastCase3}
\end{empheq}
This involves the Hankel transform of a sum of products of two correlation functions, multiplied by the power spectrum-like quantity $\mathcal{P}(k)$ from \eqq{CalPforP24}.
The coupling factor $\mathcal{M}_{3}$ from \eqq{M33def} enforces triangle conditions for $(L,\ell_2',\ell_3')$, $(L_2,\ell_1',\ell_3')$, $(L_3,\ell_1',\ell_2')$ and $(L,L_2,L_3)$, restricting the sums over $L$, $L_2$ and $L_3$ in \eqq{P24GeneralFastCase3} to be finite.

\section{Derivation of 2-4 correlations with inverse Laplacian and nontrivial angular dependence}
\label{se:P24details}

In this section we provide details for the derivation of \eqq{P24GeneralFastCase1}, which is a fast expression for contributions to $P_{24}$ that contain an inverse Laplacian with $\vp=\vq_2$ in \eqq{P24nestedStep1}. 
We explicitly show the steps for this particular case, noting that most other calculations in this paper proceed similarly in flavor but are typically less involved.

Introducing the auxiliary variable $\vq_3=\vk\!-\!\vq_2$ with a Dirac delta on the right-hand side of \eqq{P24nestedStep1}
and using \eqq{P24tadpoletmp} for the tadpole integral evaluated at $\vp=\vq_2$ gives
\begin{align}
P_{24,\mathrm{nonsep}}^{\ell_1\ell_2\ell_3\mathbf{n}s_1}(k)
& \equiv  \int\frac{\d\Omega_{\hat\vk}}{4\pi}\int_{\vq_1\vq_2} 
\,\big[\hat\vq_2\cdot\widehat{(\vk\!-\!\vq_2)}\big]^{\ell_1} 
\,\big[\hat\vq_1\cdot\widehat{(\vk\!-\!\vq_2)}\big]^{\ell_2}
\,\left(\hat{\vq}_1\cdot\hat{\vq}_2\right)^{\ell_3} 
\, \frac{q_1^{n_1}P_\lin(q_1)\,q_2^{n_2}P_\lin(q_2)\,
|\vk\!-\!\vq_2|^{n_3}
P_\lin(|\vk\!-\!\vq_2|)}{|s_1\vq_1 + \vq_2|^{2}}\non\\
& =
 (4\pi)^2\int\frac{\d\Omega_{\hat\vk}}{4\pi}\int_{\vq_2\vq_3} 
\int\d^3\vr\,
e^{i\vr\cdot(\vq_3-\vk+\vq_2)}
(\hat\vq_2\cdot\hat\vq_3)^{\ell_1} 
\,q_2^{n_2}P_\lin(q_2)
\,q_3^{n_3}P_\lin(q_3)
\non\\
\label{eq:P24helper1}
&\qquad\quad
\times
\sum_{\vl_2'}^{\ell_2}
\sum_{\vl_3'}^{\ell_3}
\sum_{\VL}^{\ell_2+\ell_3}
\alpha_{\ell_2\ell_2'}
Y^*_{\vl_2'}(\hat\vq_3)
\, 
(-1)^{L}
\mathcal{G}_{\vl_2'\VL\vl_3'}\,
Y^*_{\VL}(\hat\vq_2)\,
\mathcal{P}^{L}_{n_1,s_1}(q_2)
\,
\alpha_{\ell_3\ell_3'}
Y^*_{\vl_3'}(\hat\vq_2).
\end{align}
The angular integrals over $\hat\vk$ and $\hat\vr$ can be performed by noting that \eqq{ExpInYlms} implies
\begin{align}
  \label{eq:19}
\int\frac{\d\Omega_{\hat\vk}}{4\pi}  \int\d\Omega_{\hat\vr} \,e^{i\vr\cdot(\vq_3-\vk+\vq_2)}
= (4\pi)^2 j_0(kr)
\sum_{\VL'}^\infty
 (-1)^{L'} 
j_{L'}(q_2r)j_{L'}(q_3r) 
Y_{\VL'}(\hat\vq_2)Y^*_{\VL'}(\hat\vq_3).
\end{align}
This gives
\begin{align}
 P_{24,\mathrm{nonsep}}^{\ell_1\ell_2\ell_3\mathbf{n}s_1}(k)
& =
(4\pi)^5
\int_0^\infty\d r\,r^2\,j_0(kr)
\int_{\vq_2\vq_3}
\sum_{\vl_1'}^{\ell_1} \alpha_{\ell_1\ell_1'}
Y_{\vl_1'}(\hat\vq_2)Y^*_{\vl_1'}(\hat\vq_3)
\,q_2^{n_2}P_\lin(q_2)
\,q_3^{n_3}P_\lin(q_3)
\non\\
&
\qquad
\times\sum_{\VL'}^\infty
(-1)^{L'} 
j_{L'}(q_2r)j_{L'}(q_3r) 
Y_{\VL'}(\hat\vq_2)Y^*_{\VL'}(\hat\vq_3)
\non\\
\label{eq:P24helper2}
&
\qquad
\times\sum_{\vl_2'}^{\ell_2}
\sum_{\vl_3'}^{\ell_3}
\sum_{\VL}^{\ell_2+\ell_3}
\alpha_{\ell_2\ell_2'}
Y^*_{\vl_2'}(\hat\vq_3)
\, 
(-1)^{L}
\mathcal{G}_{\vl_2'\VL\vl_3'}\,
Y^*_{\VL}(\hat\vq_2)\,
\mathcal{P}^{L}_{n_1,s_1}(q_2)
\,
\alpha_{\ell_3\ell_3'}
Y^*_{\vl_3'}(\hat\vq_2),
\end{align}
where we also expressed $(\hat\vq_2\cdot\hat\vq_3)^{\ell_1}$ in the first line in terms of spherical harmonics using \eqq{ScalProdInYlms}.
The integral over $\hat\vq_3$ is a Gaunt integral \eq{GauntDef}, giving $\mathcal{G}_{\vl_1'\VL'\vl_2'}$.
The integral over $\hat\vq_2$ follows from \eqq{Int4Ylms}, giving 
$\sum_{\VL''}\mathcal{G}_{\vl_1'\VL'\VL''}
\mathcal{G}_{\VL''\VL\vl_3'}$. 
Using an orthogonality relation for Wigner 3-j symbols, \eqq{ThreeJOrthoSumM1M2}, the sums over $m_1'$ and $M'$ simplify to
\begin{align}
  \label{eq:8}
\sum_{m_1'M'}\mathcal{G}_{\vl_1'\VL'\vl_2'}\mathcal{G}_{\vl_1'\VL'\VL''}
=
\frac{1}{2l_2'+1}\delta_{\ell_2'L''}\delta_{m_2'M''}(\mathcal{H}_{\ell_1'L'\ell_2'})^2.
\end{align}
The same applies to the sum over $M$ and $m_3'$:
\begin{align}
  \label{eq:22}
  \sum_{Mm_3'} \mathcal{G}_{\vl_2'\VL\vl_3'}\mathcal{G}_{\VL''\VL\vl_3'}
= \frac{1}{2l_2'+1}\delta_{\ell_2'L''}\delta_{m_2'M''}(\mathcal{H}_{\ell_2'L\ell_3'})^2.
\end{align}
The sum over $m_2'$ then gives $2l_2'+1$.
The $\mathcal{H}$ factors defined in \eqq{GauntIso} contain 3-j symbols that restrict the sums over $L$ and $L'$ to be finite.
The integral over $q_3$ gives $\xi^{L'}_{n_3}(r)$.
We thus arrive at
\begin{align}
P_{24,\mathrm{nonsep}}^{\ell_1\ell_2\ell_3\mathbf{n}s_1}(k)
& =
4\pi
\int_0^\infty \d r\,r^2\,j_0(kr)
\sum_{L=0}^{\ell_2+\ell_3} 
\sum_{L'=0}^{\ell_1+\ell_2} 
\xi^{L'}_{n_3}(r)
\int\frac{\d q_2}{2\pi^2}\,
 q_2^{2+n_2}
j_{L'}(q_2r)
P_\lin(q_2)
\,\mathcal{P}^L_{n_1,s_1}(q_2)
\non\\
&\qquad\quad
\times (-1)^{L+L'}
\sum_{\ell_1'=0}^{\ell_1}
\sum_{\ell_2'=0}^{\ell_2}
\sum_{\ell_3'=0}^{\ell_3}
\alpha_{\ell_1\ell_1'}\alpha_{\ell_2\ell_2'}\alpha_{\ell_3\ell_3'}\,
\frac{(4\pi)^2}{2\ell_2'+1}
(\mathcal{H}_{\ell_1'L'\ell_2'}
\mathcal{H}_{\ell_2'L\ell_3'})^2.
\end{align}
This agrees with \eqq{P24GeneralFastCase1} above.

\section{Derivation of 3-3 correlations with inverse Laplacian}

\subsection{Simple example}
\label{se:P33SimpleDerivAppdx}
In this section we derive \eqq{P33SimpleInvDenomi}, which is a simple
example of a 3-3 
correlation with inverse Laplacian. 

Introducing $\vq_3\equiv \vk-\vq_1-\vq_2$ with a Dirac delta on the
left-hand side of \eqq{P33SimpleInvDenomi}, decomposing it in plane
waves, and performing the integral over $\vq_3$ gives
\begin{align}
  \label{eq:34}
  & \int\frac{\d\Omega_{\hat\vk}}{4\pi}\int_{\vq_1\vq_2} \frac{P_\lin(q_1)P_\lin(q_2)P_\lin(|\vk\!-\!\vq_1\!-\!\vq_2|)}{|\vq_1+\vq_2|^2}
=\int\frac{\d\Omega_{\hat\vk}}{4\pi}\int\d^3\vr\,e^{-i\vk\cdot\vr}\,
\xi^0_0(r)\,\mathbb{T}(r),
\end{align}
where we defined the 4-point like quantity $\mathbb{T}(r)$ by
\begin{align}
  \label{eq:2}
  \mathbb{T}(r) \equiv \int_{\vq_1\vq_2}
\frac{P_\lin(q_1)P_\lin(q_2)}{|\vq_1+\vq_2|^2}
\,e^{i\vq_1\cdot\vr}\,e^{i\vq_2\cdot\vr}.
\end{align}
To simplify $\mathbb{T}(r)$, we introduce $\vq_4\equiv \vq_1+\vq_2$ with
a Dirac delta and perform the integrals over $\vq_1$ and $\vq_2$ to
get
\begin{align}
  \label{eq:bbTstep2}
  \mathbb{T}(r) = \int_{\vq_4}\int\d^3\vr'\,
\frac{\xi^0_0(|\vr-\vr'|)\,\xi^0_0(|\vr-\vr'|)}{q_4^2}\,e^{i\vq_4\cdot\vr'}.
\end{align}
The integral over $\vr'$ has the form of a convolution. To solve this,
we introduce $\vr''\equiv \vr-\vr'$ with a Dirac delta, decompose this
in plane waves, and integrate over $\vr'$ to get 
(this is equivalent
to changing integration variables $\vr'\rightarrow \vr-\vr'$)
\begin{align}
  \label{eq:36}
  \mathbb{T}(r) = \int_{\vq_4}e^{i\vq_4\cdot\vr}\int\d^3\vr''\,\frac{\xi^0_0(r'')\xi^0_0(r'')}{q_4^2}\,e^{-i\vq_4\cdot\vr''}.
\end{align}
Using \eqq{IntegrateExpOverAngle}, the integral over $\hat\vr''$ gives
$4\pi j_0(q_4r'')$. Then, the integral over $\hat\vq_4$ gives $4\pi
j_0(q_4r)$. We are thus left with
\begin{align}
  \label{eq:bbTSimpleAppdx}
  \mathbb{T}(r)  
& = \int_0^\infty\frac{\d q_4}{2\pi^2}\,j_0(q_4r)
\int_0^\infty\d r''(r'')^2j_0(q_4r'')\xi^0_0(r'')\xi^0_0(r''),
\end{align}
which agrees with \eqq{bbTSimple} in the main text.

As an aside, we note that an alternative simplification of
$\mathbb{T}(r)$ follows by first integrating over $\vq_4$ in \eqq{bbTstep2},
\begin{align}
  \label{eq:14}
\mathbb{T}(r) = 
\int\d^3 \vr'\,\frac{\xi^0_0(|\vr-\vr'|)\,\xi^0_0(|\vr-\vr'|)}{4\pi r'},
\end{align}
but this 3D convolution integral does not seem suitable for fast numerical evaluation.

\subsection{General case}
\label{se:P33DerivationAppendix}

In this section we derive the right-hand side of
\eqq{P33withDenomiFast}, which allows for fast evaluation of
general 3-3 correlations with inverse Laplacian and nontrivial angular
structure in the integrand. 
The derivation proceeds similarly to the simpler example above, the
only addition being the nontrivial angular structure of the integrand,
which is taken care of by expansions in spherical harmonics.

In detail, \eqq{P33withDenomiFast} can be derived as follows.
Introducing  $\vq_4\equiv \vq_1+\vq_2$ with a second Dirac delta on the left-hand side of \eqq{P33withDenomiFast} and expanding both Dirac deltas in plane waves yields
\begin{align}
& P_{33,\mathrm{nonsep}}^{\ell_1\ell_2\ell_3,\vn}(k)
\equiv
k^{n_0}
\int\frac{\d\Omega_{\hat\vk}}{4\pi}
\int_{\vq_1\vq_2\vq_3}(2\pi)^3\delta_D(\vk-\vq_1-\vq_2-\vq_3) \frac{(\hat\vq_1\cdot\hat\vq_2)^{\ell_3}
(\hat\vq_2\cdot\hat\vq_3)^{\ell_1}
(\hat\vq_1\cdot\hat\vq_3)^{\ell_2}
}{|\vq_1+\vq_2|^{n_4}}
\prod_{i=1}^3 q_i^{n_i}P_\lin(q_i)
\\
&\;  =  
\int\frac{\d\Omega_{\hat\vk}}{4\pi}
\int\d^3\vr\int\d^3\vr' e^{-i\vk\cdot\vr}\int_{\vq_1\cdots \vq_4} e^{i\vq_4\cdot\vr'} e^{i\vq_1\cdot(\vr-\vr')}e^{i\vq_2\cdot(\vr-\vr')} e^{i\vq_3\cdot\vr} \frac{k^{n_0}}{q_4^{n_4}}
(\hat\vq_1\cdot\hat\vq_2)^{\ell_3}
(\hat\vq_2\cdot\hat\vq_3)^{\ell_1}
(\hat\vq_1\cdot\hat\vq_3)^{\ell_2}
\prod_{i=1}^3 q_i^{n_i}P_\lin(q_i).
\end{align}
Introducing $\vr''\equiv \vr-\vr'$ with another Dirac delta and expanding it in plane waves gives
\begin{align}
 P_{33,\mathrm{nonsep}}^{\ell_1\ell_2\ell_3,\vn}(k) \;=\;& 
\int\frac{\d\Omega_{\hat\vk}}{4\pi}
\int\d^3\vr \int \d^3\vr'\int\d^3\vr''\,\int_{\vq}\int_{\vq_1\cdots \vq_4} e^{i\vq\cdot(\vr''-\vr+\vr')}  e^{-i\vk\cdot\vr} e^{i\vq_4\cdot\vr'} e^{i\vq_1\cdot\vr''}e^{i\vq_2\cdot\vr''} e^{i\vq_3\cdot\vr}
\non\\
&\times  \frac{k^{n_0}}{q_4^{n_4}}\,(\hat\vq_1\cdot\hat\vq_2)^{\ell_3}
(\hat\vq_2\cdot\hat\vq_3)^{\ell_1}
(\hat\vq_1\cdot\hat\vq_3)^{\ell_2}
\prod_{i=1}^3 q_i^{n_i}P_\lin(q_i).
\end{align}
Using \eqq{IntegrateExpOverAngle}, the integral over $\hat\vk$ gives $j_0(kr)$, and the integral over $\hat\vq_4$ gives $4\pi j_0(q_4r')$. Then, the integral over $\hat\vr'$ yields $4\pi j_0(qr')$.
Next we expand the remaining plane waves and the scalar products between wavevectors in spherical harmonics (using \eqq{ScalProdInYlms} for the latter). This leads to two spherical harmonics with argument $\hat\vq$, so that the integral over $\hat\vq$ gives a Kronecker delta. The same happens for the integral over $\hat\vr$.  Additionally, there are three spherical harmonics with argument $\hat\vq_1$, so that integrating over $\hat\vq_1$ gives a Gaunt integral \eq{GauntDef}. The same happens for integrals over $\hat\vq_2$, $\hat\vq_3$ and $\hat\vr''$.
The integral over $r'$ follows from the closure relation for spherical Bessel functions, enforcing $q_4=q$, and yielding the intermediate result
 \begin{align}
    P_{33,\mathrm{nonsep}}^{\ell_1\ell_2\ell_3,\vn}(k) =\;& k^{n_0} \frac{(4\pi)^{10}}{[(2\pi)^3]^5}\frac{\pi}{2} \int_0^\infty \d r\,\d r''\d q\,\d q_1\,\d q_2\,\d q_3\, r^2 (r'')^2 \, q^{2-n_4} 
\left[\prod\limits_{i=1}^3 q_i^{2+n_i}P_\lin(q_i)\right]\non\\
&\times
\sum\limits_{L_{1,2,3}\ell'_{1,2,3}} 
i^{L_1+L_2+L_3} 
(-1)^{\ell_1'+\ell_2'+\ell_3'}
\,\alpha_{\ell_1\ell_1'}\alpha_{\ell_2\ell_2'}\alpha_{\ell_3\ell_3'}
\,j_0(kr) j_{L_3}(qr)j_{L_3}(qr'') j_{L_1}(q_1r'') j_{L_2}(q_2r'') j_{L_3}(q_3r)
\non\\
\label{eq:P33fullintermed}
&\times \sum\limits_{M_{1,2,3}m'_{1,2,3}} (-1)^{m_1'+m_2'+m_3'}
\mathcal{G}_{L_1\ell_2'\ell_3'}^{M_1m_2'm_3'}
\mathcal{G}_{L_2\ell_1'\ell_3'}^{M_2m_1'-m_3'}
\mathcal{G}_{L_3\ell_1'\ell_2'}^{M_3-m_1'-m_2'}
\mathcal{G}_{L_1L_2L_3}^{M_1M_2M_3}.
 \end{align}
The last sum of four Gaunt integrals over $M_i$ and $m'_i$ gives a 6-j symbol \eq{modified6j}.
Conveniently arranging the integration order then allows to write the $P_{33}$ integral as a sum over 1D Hankel transforms as in \eqq{P33withDenomiFast} in the main text.

Some comments regarding the derivation are in order.
In this section we only consider a quadratic denominator, $n_4=2$, corresponding to a single inverse Laplacian, but the calculation above formally works for any $n_4$, which is why we gave the result for arbitrary $n_4$ (noting though that some integrals may diverge for $n_4\ne 2$).
In the special case without denominator, $n_4=0$, the integral over $q$ in \eqq{XdefP33} yields a Dirac delta enforcing $r=r''$,
so that we recover \eqq{P33simple1}.
As an alternative way to simplify the integral on the left-hand side of \eqq{P33withDenomiFast} one could 
split the derivative with respect to $\vr$ into radial and angular parts, similarly to e.g.~\cite{kendrickTrispectrum1502}. This is convenient for $n_4=2$ but gets likely more complicated for larger $n_4$, whereas the results above apply to general $n_4$.

\end{widetext}

\section{Specialization to scaling universe with perfect power law initial power spectrum}
\label{se:ScalingUni}
The DM power spectrum in a $\Lambda$CDM cosmology is scale-dependent and includes features like baryonic acoustic oscillations. 
Therefore, integrals over this power spectrum or Hankel transforms must be performed \emph{numerically}, e.g.~using \textsf{FFTLog} \cite{hamiltonfftlog}. 
This is the primary use case we envision for our method, because it allows evaluating 2-loop corrections to the matter power spectrum in an extremely fast way for arbitrary shapes of the initial linear power spectrum. 
In this section we specialize the general results from the rest of the paper to a simpler special case in which the transforms can actually be performed \emph{analytically}, exploiting the fact that the Hankel transform of a power law is again a power law.
This may be useful for validating numerical implementations, but we stress again that it is not needed for our method which applies to arbitrary linear power spectrum shapes.

For scaling universes the DM power spectrum is assumed to have a power-law shape, 
\begin{align}
  \label{eq:9}
 P_\lin(k)=\left(\frac{k}{k_0}\right)^N,
\end{align}
with some slope $N$ and pivot scale $k_0$. 
The linear correlation function $\xi^\ell_n$ then reduces to
\begin{align}
  \label{eq:29}
  \left[\xi^l_n(r)\right]_\text{scal.uni.} & \;=\;
\frac{1}{k_0^N} \int_0^\infty \frac{\d q}{2\pi^2}\,
q^{2+n+N}\,j_l(qr)
 \non\\
 & \;=\;
  \frac{\Xi^l_{n+N}(r)}{k_0^N},
\end{align}
where the integral over $q$ is a Hankel transform of a power law.
This is again a power law 
\begin{align}
  \label{eq:11}
\Xi^l_{n+N}(r) 
&= \frac{2^{n+N}}{\pi\sqrt{\pi}}\frac{\Gamma\left(\frac{l+n+N+3}{2}\right)}{\Gamma\left(\frac{l-n-N}{2}\right)}\frac{1}{r^{3+n+N}}
\end{align}
if $n+N<-1$, $n+N+l>-3$ and $r>0$ \cite{MarcelZvonimirPat1603,NISTDLMF,Olver:2010:NHMF,Watson66}.
In the rest of this section we use this result to specialize the fast expressions in the main text of the paper to scaling universes, obtaining analytical solutions for all 2-loop integrals in scaling universes.

\subsection{1-5 correlations in a scaling universe}

For example, in a scaling universe, the fast expression for 1-5 correlations given by the right-hand side of \eqq{P15GeneralFast} reduces to a simple power law in $k$
(assuming $s_0=s_1=s_2=1$)
\begin{align}
  \label{eq:P15ScalUni}
\left[\text{\eqq{P15GeneralFast}}\right]_\text{scal.uni.} = 
  c_{15}\,\frac{k^{n_0+N}}{k_0^N}\,k^{4+2N+n_1+n_2},
\end{align}
where the proportionality constant is
\begin{align}
  \label{eq:1}
  c_{15} = \, &  \frac{\sqrt{\pi}}{(4\pi)^3(k_0)^{2N}}
\sum_{L_0=0}^{\ell_1+\ell_2}
\frac{\Gamma\left(\frac{L_0-4-2N-n_1-n_2}{2}\right)}{\Gamma\left(\frac{L_0+7+2N+n_1+n_2}{2}\right)}\non\\
&\times\sum_{L_1=0}^{\ell_0+\ell_1}\sum_{L_2=0}^{\ell_0+\ell_2}
\mathcal{M}_{3}(\ell_0,\ell_1,\ell_2;L_0,L_2,L_1)\non\\
&\times\prod_{i=1}^2\frac{\Gamma\left(\frac{L_i+n_i+N+3}{2}\right)}{\Gamma\left(\frac{L_i-n_i-N}{2}\right)}.
\end{align}
For $N=-2.6$ and $\ell_i=n_i=0$, we validated \eqq{P15ScalUni} numerically by brute-force integrating the left-hand side using the Monte-Carlo integration library \textsf{Cuba} \cite{cuba}.

\subsection{2-4 correlations in a scaling universe}

We can also simplify the fully general 2-4 correlations of \eqq{P24GeneralFastCase1} in a perfectly scaling universe. 
To see this, note that the transformed correlation $\mathcal{P}(p)$ of \eqq{CalPforP24} becomes a power law in a scaling universe,
\begin{align}
  \label{eq:31}
 \left[\mathcal{P}^L_{n_1,1}(p)\right]_\text{scal.uni.} & =
\frac{1}{8\pi k_0^N}\,
\frac{\Gamma\left(\frac{L+n_1+N+3}{2}\right)}{\Gamma\left(\frac{L-n_1-N}{2}\right)}\non\\
&\quad\;\times\frac{\Gamma\left(\frac{L-n_1-N-1}{2}\right)}{\Gamma\left(\frac{L+n_1+N+4}{2}\right)}\,
p^{1+n_1+N}.
\end{align}
Then the 4-point-like correlation $\mathcal{T}(r)$ in \eqq{Tau24General} also becomes a power law,
\begin{align}
  \label{eq:32}
  \left[\mathcal{T}^{1,n_1n_2}_{LL'}(r)\right]_\text{scal.uni.} 
&= \frac{2^{1+n_1+n_2+2N}\left[\mathcal{P}^L_{n_1,1}(1)\right]_\text{scal.uni.}}{\pi\sqrt{\pi}\,k_0^N\,r^{4+n_1+n_2+2N}} \non\\
&\quad\;\times
\frac{\Gamma\left(\frac{L'+n_1+n_2+2N+4}{2}\right)}{\Gamma\left(\frac{L'-n_1-n_2-2N-1}{2}\right)}.
\end{align}
Therefore the right-hand side of the 2-4 integral in \eqq{P24GeneralFastCase1} also turns into a simple power law in $k$ (assuming $s_1=1$),
\begin{align}
  \label{eq:P24ScalUni1a}
  \left[\text{Eq.~}\eq{P24GeneralFastCase1}\right]_\text{scal.uni.}
= c_{24}\, k^{4+n_1+n_2+n_3+3N}
\end{align}
with proportionality constant
  \begin{align}
    \label{eq:33}
    c_{24} & = \frac{1}{2^{6+n_1+n_2+2N}k_0^N}
\frac{\Gamma\left(\frac{-4-n_1-n_2-n_3-3N}{2}\right)}
{\Gamma\left(\frac{7+n_1+n_2+n_3+3N}{2}\right)}\non\\
&\quad\times
\sum_{L=0}^{\ell_2+\ell_3}
\sum_{L'=0}^{\ell_1+\ell_2}
\mathcal{M}_{2}(\ell_1,\ell_2,\ell_3;L,L')\non\\
&\quad\times
\left[\mathcal{T}^{1,n_1n_2}_{LL'}(1)\right]_\text{scal.uni.}
\frac{\Gamma\left(\frac{L'+n_3+N+3}{2}\right)}
{\Gamma\left(\frac{L'-n_3-N}{2}\right)}.
  \end{align}

The other case of general 2-4 correlations is given by \eqq{P24GeneralFastCase3}. This takes the same power law form in a scaling universe (assuming again $s_1=1$)
\begin{align}
  \label{eq:P24ScalUni2a}
    \left[\text{Eq.~}\eq{P24GeneralFastCase3}\right]_\text{scal.uni.}
= c'_{24}\, k^{4+n_1+n_2+n_3+3N}
\end{align}
but the proportionality constant is now
\begin{align}
  \label{eq:35}
  c'_{24}  = &\; 
\sum_{L=0}^{\ell_2+\ell_3}
\frac{\left[\mathcal{P}^L_{n_1,1}(1)\right]_\text{scal.uni.}}
{8\pi\sqrt{\pi}\,k_0^{2N}}
\frac{\Gamma\left(\frac{L-n_2-n_3-2N-3}{2}\right)}
{\Gamma\left(\frac{L+n_2+n_3+2N+6}{2}\right)}\non\\
&\quad\times \sum_{L_2=0}^{\ell_1+\ell_3}\sum_{L_3=0}^{\ell_1+\ell_2}
\mathcal{M}_{3}(\ell_1,\ell_2,\ell_3;L,L_2,L_3)\non\\
&\quad\times
\frac{\Gamma\left(\frac{L_2+n_2+N+3}{2}\right)}
{\Gamma\left(\frac{L_2-n_2-N}{2}\right)}
\frac{\Gamma\left(\frac{L_3+n_3+N+3}{2}\right)}
{\Gamma\left(\frac{L_3-n_3-N}{2}\right)}.
\end{align}

For $N=-2.6$ and $n_i=\ell_i=0$, Eqs.~\eq{P24ScalUni1a} and
\eq{P24ScalUni2a} are consistent with results obtained by Monte-Carlo
integrating the left-hand side.

\subsection{3-3 correlations in a scaling universe}

The fast expression for 3-3 correlations without inverse Laplacians given by the right-hand side of \eqq{P33simple1} also reduce to a
simple power law in $k$ for scaling universes,
\begin{align}
  \label{eq:3}
\left[\text{\eqq{P33simple1}}\right]_\text{scal.uni.} =  
c_{33} \, k^{6+n_1+n_2+n_3+3N},
\end{align}
where the proportionality constant is
\begin{align}
  c_{33} =\, &  \frac{1}{(4\pi)^3 (k_0)^{3N}}
\frac{\Gamma\left(\frac{-n_1-n_2-n_3-3N-6}{2}\right)}{\Gamma\left(\frac{n_1+n_2+n_3+3N+9}{2}\right)}\non\\
& \times\sum_{L_1=0}^{\ell_2+\ell_3}\sum_{L_2=0}^{\ell_1+\ell_3}\sum_{L_3=0}^{\ell_1+\ell_2}
\mathcal{M}_{3}(\ell_1,\ell_2,\ell_3;L_1,L_2,L_3)\non\\
&\times\prod_{i=1}^3\frac{\Gamma\left(\frac{L_i+n_i+N+3}{2}\right)}{\Gamma\left(\frac{L_i-n_i-N}{2}\right)}.
  \label{eq:c33constant}
\end{align}
We numerically validated this both for scaling universes and for a realistic linear input power spectrum for $\ell_i=n_i=0$.

More general 3-3 correlations with inverse Laplacians are given in \eqq{P33withDenomiFast}. For a scaling universe, the 4-point like-quantity $\mathcal{T}(r)$ from \eqq{XdefP33} becomes a power law,
\begin{align}
  \label{eq:bla}
&  \left[\mathbb{T}^{\ell_1\ell_2\ell_3,n_1n_2n_4}_{L_3}(r)\right]_\text{scal.uni.}
=
\frac{2^{n_1+n_2-n_4+2N-2}}{\pi^4\,k_0^{2N}\,r^{n_1+n_2-n_4+2N+6}}\non\\
&\quad\times
\frac{\Gamma\left(\frac{L_3+n_1+n_2-n_4+2N+6}{2}\right)}
{\Gamma\left(\frac{L_3-n_1-n_2+n_4-2N-3}{2}\right)}
\frac{\Gamma\left(\frac{L_3-n_1-n_2-2N-3}{2}\right)}
{\Gamma\left(\frac{L_3+n_1+n_2+2N+6}{2}\right)}\non\\
&\quad\times\sum_{L_1=|\ell_2-\ell_3|}^{\ell_2+\ell_3}\sum_{L_2=|\ell_1-\ell_3|}^{\ell_1+\ell_3}\mathcal{M}_{3}(\ell_1,\ell_2,\ell_3;L_1,L_2,L_3)\non\\
&\quad\times
\frac{\Gamma\left(\frac{L_1+n_1+N+3}{2}\right)}
{\Gamma\left(\frac{L_1-n_1-N}{2}\right)}
\frac{\Gamma\left(\frac{L_2+n_2+N+3}{2}\right)}
{\Gamma\left(\frac{L_2-n_2-N}{2}\right)}.
\end{align}
The right-hand side of \eqq{P33withDenomiFast} therefore becomes
\begin{align}
  \label{eq:38}
  \left[\text{\eqq{P33withDenomiFast}}\right]_\text{scal.uni.} =  
c'_{33}\,k^{6+n_1+n_2+n_3-n_4+3N},
\end{align}
with proportionality constant
 \begin{align}
   \label{eq:10}
&c'_{33} =
\frac{1}{(2\pi)^3(k_0)^{3N}}
\frac{\Gamma\left(\frac{-n_1-n_2-n_3+n_4-3N-6}{2}\right)}
{\Gamma\left(\frac{n_1+n_2+n_3-n_4+3N+9}{2}\right)}
\sum_{L_3=|\ell_1\!-\!\ell_2|}^{\ell_1+\ell_2}
  \non\\
&\;\;
\frac{\Gamma\left(\frac{L_3+n_1+n_2-n_4+2N+6}{2}\right)}
{\Gamma\left(\frac{L_3-n_1-n_2+n_4-2N-3}{2}\right)}
\frac{\Gamma\left(\frac{L_3-n_1-n_2-2N-3}{2}\right)}
{\Gamma\left(\frac{L_3+n_1+n_2+2N+6}{2}\right)}
\sum_{L_1=|\ell_2\!-\!\ell_3|}^{\ell_2+\ell_3}
\non\\
&\;\;
\sum_{L_2=|\ell_1\!-\!\ell_3|}^{\ell_1+\ell_3}
\mathcal{M}_{3}(\ell_1,\ell_2,\ell_3;L_1,L_2,L_3)
\prod_{i=1}^3\frac{\Gamma\left(\frac{L_i+n_i+N+3}{2}\right)}{\Gamma\left(\frac{L_i-n_i-N}{2}\right)}.
\end{align}
Numerically validating this result is unfortunately not straightforward because
the brute force integration of the left-hand side seems nontrivial for
scaling universes. Nevertheless, the predicted scaling with $k$ does
seem consistent with brute force integration if we choose $N=-2.1$.

\section{Useful mathematical identities}

For convenience we list some standard mathematical identities that we used throughout this paper (also see Appendix C in \cite{MarcelZvonimirPat1603}).

\subsection{Expansions}
Some of the most frequently used relations in our paper are the expansion of a Dirac delta in plane waves,
\begin{align}
  \label{eq:DiracDeltaInPlaneWaves}
  (2\pi)^3\delta_D(\vq) = \int\d^3 \vr\,e^{i\vq\cdot\vr},
\end{align}
the expansion of plane waves in spherical harmonics,
\begin{align}
  \label{eq:ExpInYlms}
  e^{\pm i a\vk\cdot\vr} = 4\pi \sum_{l=0}^\infty\sum_{m=-l}^l (\pm i\, \sgn(a))^l\, j_l(|a|kr)\,Y_{lm}(\hat\vk)Y^*_{lm}(\hat\vr),
\end{align}
and the decomposition of scalar products between wavevectors into spherical harmonics,
\begin{align}
  \label{eq:ScalProdInYlms}
(\hat\vx\cdot\hat{\vy})^\ell=4\pi\sum_{\ell'=0}^\ell\sum_{m'=-\ell'}^{\ell'}
\alpha_{\ell\ell'}Y_{\ell'm'}(\hat{\vx})Y^*_{\ell'm'}(\hat{\vy}),
\end{align}
where $\alpha_{\ell\ell'}$ coefficients are given by \eqq{alphas}.

\subsection{Angular integrals and Wigner 3-j symbols}
The integral over three spherical harmonics is a Gaunt integral that contains Wigner 3-j symbols,
\begin{align}
\int\d\Omega_{\hat{\vq}}\, Y_{\vl_1}(\hat{\vq})&Y_{\vl_2}(\hat{\vq})Y_{\vl_3}(\hat{\vq})
= 
\mathcal{G}_{\vl_1 \vl_2\vl_3}\non\\
&\qquad = \mathcal{H}_{\ell_1\ell_2\ell_3}
\bigg(\begin{matrix}
  \ell_1 & \ell_2 & \ell_3\\
   m_1 & m_2 & m_3
\end{matrix}\bigg),
\label{eq:GauntDef}
\end{align}
where the isotropic part is
\begin{align}
  \label{eq:GauntIso}
  \mathcal{H}_{\ell_1\ell_2\ell_3} \equiv \sqrt{\frac{(2\ell_1+1)(2\ell_2+1)(2\ell_3+1)}{4\pi}}
\bigg(\begin{matrix}
  \ell_1 & \ell_2 & \ell_3\\
  0 & 0 & 0
\end{matrix}\bigg).
\end{align}
The indices must satisfy $m_1+m_2+m_3=0$, $|\ell_2-\ell_3|\le \ell_1 \le \ell_2+\ell_3$ and permutations, and $\ell_1+\ell_2+\ell_3$ must be even.
The Gaunt coefficients \eq{GauntDef} represent the coefficients that arise when decomposing the product of two spherical harmonics in terms of a third one, i.e.
\begin{align}
  \label{eq:TwoYsInTermsOfOne}
  Y_{\vl_1}(\hat\vq)\,Y_{\vl_2}(\hat\vq) \;=\; \sum_{L=|\ell_1-\ell_2|}^{\ell_1+\ell_2}\sum_{M=-L}^{L} \mathcal{G}_{\vl_1\vl_2\VL}\,Y^*_{\VL}(\hat\vq).
\end{align}
The integral over four spherical harmonics is therefore
\begin{align}
  \label{eq:Int4Ylms}
  \int\d\Omega_{\hat\vq}\, Y_{\vl_1}(\hat\vq) Y_{\vl_2}(\hat\vq)
Y^*_{\vl_3}(\hat\vq)Y^*_{\vl_4}(\hat\vq)
= \sum_{\VL} \mathcal{G}_{\vl_1\vl_2\VL}\,\mathcal{G}_{\VL\vl_3\vl_4}.
\end{align}

The Wigner 3-j symbols satisfy the following orthogonality relation:
\begin{align}
  \label{eq:ThreeJOrthoSumM1M2}
  \sum_{m_1m_2}
\bigg(\begin{matrix}
  \ell_1 & \ell_2 & \ell \\
  m_1 & m_2 & m
\end{matrix}\bigg)
\bigg(\begin{matrix}
  \ell_1 & \ell_2 & \ell' \\
  m_1 & m_2 & m'
\end{matrix}\bigg)
=
\frac{1}{2\ell+1}\delta_{\ell\ell'}\delta_{mm'}.
\end{align}
Some other relations used in our paper are
\begin{align}
  \label{eq:Sum3JOverM}
  \sum_m (-1)^m 
\bigg(\begin{matrix}
  \ell & \ell & L \\
  m & -m & 0
\end{matrix}\bigg)
=
(-1)^\ell \sqrt{2\ell+1}\delta_{L0}
\end{align}
and
\begin{align}
  \label{eq:3JSimpleCase}
  \bigg(\begin{matrix}
  \ell & \ell & 0 \\
  m & -m & 0
\end{matrix}\bigg)
= \frac{(-1)^{\ell-m}}{\sqrt{2\ell+1}}.
\end{align}
Eqs.~\eq{TwoYsInTermsOfOne}-\eq{3JSimpleCase} can be used 
to perform the angular integral over the product
of four Legendre polynomials
with the same argument
\begin{align}
&  \int\d\Omega_{\hat\vr}\,
\mathsf{P}_{\ell_1}(\hat\vk\cdot\hat\vr)
\mathsf{P}_{\ell_2}(\hat\vk\cdot\hat\vr)
\mathsf{P}_{\ell_3}(\hat\vk\cdot\hat\vr)
\mathsf{P}_{\ell_4}(\hat\vk\cdot\hat\vr) \non\\
  &\quad=4\pi \sum_{L} (2L+1)
  \left(\begin{matrix}
  L  & \ell_1 & \ell_2\\
  0 & 0 & 0
\end{matrix}\right)^2
  \left(\begin{matrix}
  L  & \ell_3 & \ell_4\\
  0 & 0 & 0
\end{matrix}\right)^2,
\label{eq:IntegrateFourLegendres}
\end{align}
where the sum over $L$ is restricted by triangle conditions.

The angular part of the 3D Fourier transform of a function $f(r)$ that depends only on radius is
\begin{align}
  \label{eq:IntegrateExpOverAngle}
  \int\d\Omega_{\hat\vq}\,  e^{\pm i\vq\cdot\vr}\, f(r)
\,=\,4\pi\,j_0(qr)\,f(r).
\end{align}
The Fourier transform of the inverse Laplacian is (e.g.~\cite{MarcelZvonimirPat1603})
\begin{align}
  \label{eq:FTofInvLaplacian}
\Xi^0_{-2}(r) \equiv \int_{\vq}e^{-i\vq\cdot\vr}\frac{1}{q^2}
=   \int_0^\infty \frac{\d q}{2\pi^2} j_0(qr)=\frac{1}{4\pi r}.
\end{align}

\subsection{Wigner 6-j symbol}
We sometimes use a rescaled 6-j symbol defined by
\begin{align}
&&\bigg\{\begin{matrix}
j_1 &  j_2 & j_3  \\
j_4 & j_5 & j_6 
\end{matrix}\bigg\}'
 \equiv 
(4\pi)^2\,
i^{j_1+j_2+j_3} (-1)^{j_4+j_5+j_6}\,
\mathcal{H}_{j_1j_2j_3}\quad\;\;
\non\\
&&\quad\;\times\mathcal{H}_{j_1j_5j_6}
\mathcal{H}_{j_2j_4j_6}\mathcal{H}_{j_3j_4j_5}\;
\bigg\{\begin{matrix}
j_1 &  j_2 & j_3  \\
j_4 & j_5 & j_6 
\end{matrix}\bigg\}
\quad
\non\\
&&
  \label{eq:modified6j}
 = 
i^{j_1+j_2+j_3}  (-1)^{j_4+j_5+j_6}\,
\bigg(\begin{matrix}
j_1 &  j_2 & j_3  \\
0 & 0 & 0 
\end{matrix}\bigg)  
\bigg(\begin{matrix}
j_1 &  j_5 & j_6  \\
0 & 0 & 0 
\end{matrix}\bigg)  
\quad\;\;
\non\\
&&
\quad\times
\bigg(\begin{matrix}
j_2 &  j_4 & j_6  \\
0 & 0 & 0 
\end{matrix}\bigg)  
\bigg(\begin{matrix}
j_3 &  j_4 & j_5  \\
0 & 0 & 0 
\end{matrix}\bigg)  
\bigg\{\begin{matrix}
j_1 &  j_2 & j_3  \\
j_4 & j_5 & j_6 
\end{matrix}\bigg\} 
\prod_{i=1}^6 (2j_i+1).
\end{align}
This is only nonzero if triangle conditions of the form $|j_2-j_3|\le j_1\le j_2+j_3$ are satisfied for $(j_1,j_2,j_3)$, $(j_1,j_5,j_6)$, $(j_2,j_4,j_6)$ and $(j_3,j_4,j_5)$.
Additionally, $j_1+j_2+j_3$, $j_1+j_5+j_6$, $j_2+j_4+j_6$ and $j_3+j_4+j_5$ must be even.
The product of the first 3-j symbol and the 6-j symbol in \eqq{modified6j} can also be replaced by a sum over 3-j symbols using Eq.~34.5.23 of \cite{NISTDLMF,Olver:2010:NHMF}:
\begin{align}
&\bigg\{\begin{matrix}
j_1 &  j_2 & j_3  \\
j_4 & j_5 & j_6 
\end{matrix}\bigg\}'
=
i^{j_1+j_2+j_3} 
\bigg(\begin{matrix}
j_1 &  j_5 & j_6  \\
0 & 0 & 0 
\end{matrix}\bigg)  
\bigg(\begin{matrix}
j_2 &  j_4 & j_6  \\
0 & 0 & 0 
\end{matrix}\bigg)  
\non\\
&\quad\times
\bigg(\begin{matrix}
j_3 &  j_4 & j_5  \\
0 & 0 & 0 
\end{matrix}\bigg)  
\bigg[\prod_{n=1}^6 (2j_n+1)\bigg]
\sum_{m=-\mathrm{max}(j_4,j_5,j_6)}^{\mathrm{max}(j_4,j_5,j_6)} (-1)^{m} \non\\
&\quad
\times
\bigg(\begin{matrix}
j_1 &  j_5 & j_6  \\
0 & m & -m
\end{matrix}\bigg)
\bigg(\begin{matrix}
j_2 & j_4 &   j_6  \\
0 & m & -m
\end{matrix}\bigg)
\bigg(\begin{matrix}
j_3 & j_4 &  j_5   \\
0 & m & -m 
\end{matrix}\bigg).
\end{align}
Numerical evaluation is straightforward and fast, noting that we only require $j_i\lesssim 10$ because the perturbation theory kernels $F_n$ and $G_n$ involve only low-order Legendre polynomials.

\bibliography{bib_fft_loops}

\end{document}